\documentclass[preprint]{elsarticle}
\usepackage{amssymb,amsmath}
\usepackage{units}

\usepackage{graphicx}
\usepackage{morefloats}

\def\menorsim{\smash{\mathop{<}\limits_{\raise3pt\hbox{$\sim$}}}}
\def\maiorsim{\smash{\mathop{>}\limits_{\raise3pt\hbox{$\sim$}}}}

\begin{document}               
\begin{frontmatter}
\title{Characterisation of the electromagnetic component in ultra-high energy inclined air showers}

\author[kar]{I.~Vali\~no\corref{cor1}}
\ead{inesvr@gmail.com}
\author[sant]{J.~Alvarez-Mu\~niz}
\author[kar]{M. Roth}
\author[sant]{R.A.~Vazquez}
\author[sant]{E.~Zas}

\address[kar]{Karlsruhe Institute of Technology, POB 3640, D-76021
  Karlsruhe, Germany} 
\address[sant]{Instituto Galego de F\'\i sica de
  Altas Enerx\'\i as and Departamento de F\'\i sica de Part\'\i
  culas,\\ Universidade de Santiago de Compostela, Spain}

\cortext[cor1]{Corresponding author. Present address: Institut f\"ur
  Kernphysik, Karlsruhe Institute of Technology, D-76021 Karlsruhe,
  Germany. Tel: +49 (0) 724782-4978.}

\begin{abstract}
Inclined air showers -- those arriving at ground with zenith angle
with respect to the vertical $\theta > 60^{\circ}$ -- are
characterised by the dominance of the muonic component at ground which
is accompanied by an electromagnetic ``halo" produced mainly by muon
decay and muon interactions. By means of Monte Carlo simulations we
give a full characterisation of the particle densities at ground in
ultra-high energy inclined showers as a function of primary energy and
mass composition, as well as for different hadronic models assumed in
the simulations. We also investigate the effect of intrinsic
shower-to-shower fluctuations in the particle densities.
\end{abstract}

\end{frontmatter}

\section{Introduction}

Inclined air showers are conventionally defined as those arriving at
ground with zenith angles $\theta$ above $60^\circ$.  At large zenith
angles the electromagnetic~(EM) component in air showers, mainly produced
by the decay of $\pi^0$s, is largely absorbed in the greatly enhanced
atmospheric depth the shower needs to cross before reaching ground, so
that in a first approximation only the more penetrating particles such
as muons survive to ground.  Muons are accompanied by an
electromagnetic component produced mainly by muon decay in flight and
muon interactions such as bremsstrahlung, pair production and nuclear
interactions. A full characterisation of this so-called
electromagnetic ``halo" is given in this work.

The study of inclined showers is of great interest because their
detection immediately enhances the exposure of existing air shower
detectors by up to about $30\%$ with respect to that achieved with
vertical showers $(\theta<60^\circ)$, extending the field of view to
sky directions otherwise inaccessible.

Inclined showers have been detected in the past in arrays of ground
detectors such as the Haverah Park~\cite{Ave:1999cp} and
AGASA~\cite{Yoshida:2001pw} experiments. Modern detectors such as the
Pierre Auger Observatory~\cite{Abraham:2004dt} or the Telescope
Array~\cite{TA,Kawai:2008zza} can also be used to detect showers with
large zenith angles.  In particular, the Surface Detector Array (SD)
of the Pierre Auger Observatory is well suited to detect very inclined
showers at energies above about $5 \times 10^{18}$ eV, with high
efficiency and unprecedented statistical accuracy. The energy spectrum
of Ultra-High Energy Cosmic Rays (UHECR) with inclined showers was
recently measured~\cite{FacalSanLuis:2007it}. The surface detector is
an extended array of deep water Cherenkov detectors that act like
volume detectors adequate for recording particles arriving at ground
at all zenith angles.

As in all ground arrays, the distribution of the detector signals
produced by shower particles is used to estimate shower observables
such as the primary energy and to perform composition studies. As
shown in previous works~\cite{Inoue:1999cn,Ave:2000xs}, the specific
characteristics of inclined showers entail that their analysis
requires a different approach from the standard one for vertical
showers. The study of the particle densities of the electromagnetic
and muonic components at ground level becomes essential in the
reconstruction~\cite{Newton:2007qi} and analysis of events at large
zenith angles.

Inclined showers have also been studied for years for other
reasons. In the 70's, it was suggested that showers induced by
neutrinos could be identified in the background of inclined proton or
nuclei induced showers by searching for deeply penetrating inclined
showers~\cite{Berezinsky1975} which should exhibit a significant
electromagnetic component at ground. In this respect the study and
characterisation of the electromagnetic halo in inclined showers is
also of great importance.

In this paper we have used Monte Carlo generators simulating the air
shower development to obtain the particle densities induced by the
electromagnetic and muonic components of inclined air showers. 
In contrast to other works~\cite{Ave:1999cp,Cillis:2000xc} we have
performed a comprehensive characterisation of the ratio of the
electromagnetic to muonic densities as a function of distance to the
shower core for different zenith angles. We have studied the
dependences of this ratio on the primary energy, mass composition and
hadronic interaction model assumed in the simulations. We have also
studied the effect of the geomagnetic field that deviates charged
particles along their paths to the detector. As a result of this study
we give parameterisations of the average ratio of the EM and muonic
components as a function of the distance to the shower core, shower
zenith angle~($\theta$) and azimuth angle ($\zeta$) of the position at
which the particle arrives at ground with respect to the incoming
shower direction, as well as on the distance to the shower core in the
shower plane~\footnote{The shower plane is the plane transverse to the
  shower axis.}.
 
The effect of the intrinsic shower-to-shower fluctuations on the
electromagnetic and muonic contributions to the density has also been
studied. Our results are relevant for the reconstruction of inclined
showers in ultra-high energy cosmic ray detectors such as the Pierre
Auger Observatory and the Telescope Array.

\section{Simulation method}

\subsection{Extensive air shower simulations}

We have generated a library of inclined air showers induced by proton
and iron primaries using the Monte Carlo shower propagation code AIRES
2.6.0~\cite{aires} and the hadronic interaction models
QGSJET01~\cite{qgsjet} and Sibyll 2.1~\cite{Engel:1999db}. Showers
were generated with energies of 1, 10 and \unit{100} {EeV}, with
zenith angle ranging from 60$^{\circ}$ to 88$^{\circ}$ in steps of
2$^{\circ}$ and random azimuth angle (unless otherwise indicated). We
have simulated 100 showers for each combination of energy and zenith
angle. Showers were simulated with and without geomagnetic field
assuming the location of the surface detector of the Pierre Auger
Observatory~\cite{Abraham:2004dt} (ground level placed at a depth of
$876~{\rm g~cm^{-2}}$) and using a curved atmosphere based on the
Linsley's atmospheric model~\cite{Linsley}.

The number of particles that are produced in an air shower at EeV
energies and above and the computing time needed to follow all the
secondaries is typically excessively large. To avoid this problem,
current air shower simulation codes use a statistical sampling
algorithm (thinning algorithm) which allows to propagate only a small
representative fraction of the total number of
particles~\cite{Hillas:1997tf}. Statistical weights are assigned to
the sampled particles in order to compensate for the rejected
ones. Therefore, the output of the air shower simulation is a detailed
data file with weighted entries containing the particle information at
ground level.  In this work, showers were simulated with a relative
thinning level of $10^{-6}$ and a weight factor for the
electromagnetic (heavy) particles of 12 (0.14). Since the thinning
process itself introduces artificial fluctuations, for the study of
the intrinsic shower-to-shower fluctuations we have also simulated
sets showers with a thinning level of $10^{-7}$. All the particles
with kinetic energies above the following thresholds were tracked: 80~keV 
for electrons, positrons and gammas, 10~MeV for muons, 60~MeV for
mesons and 120~MeV for nucleons and nuclei.  

\begin{figure*}
\begin{center}
\includegraphics[width=0.49\textwidth]{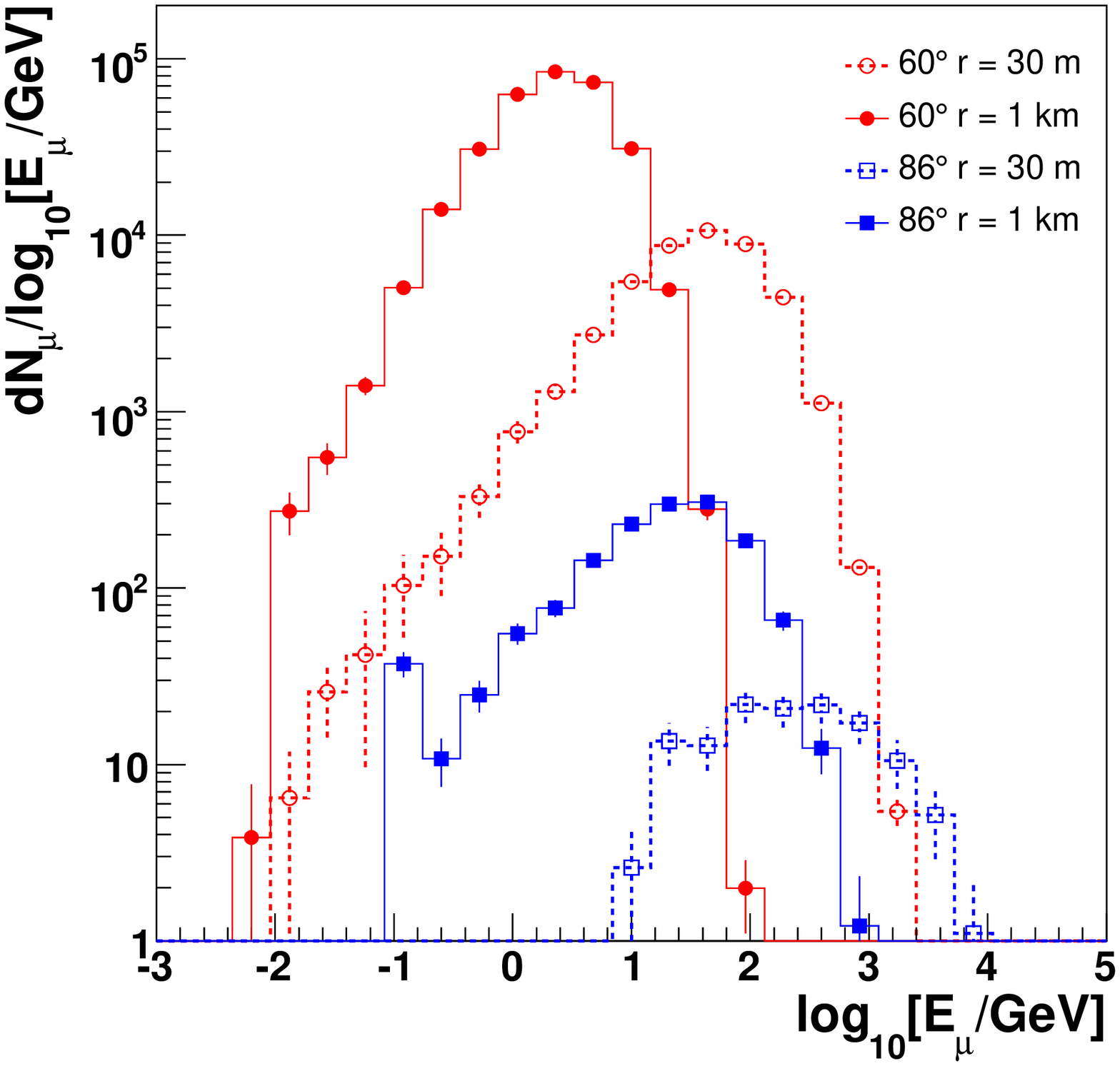}
\includegraphics[width=0.49\textwidth]{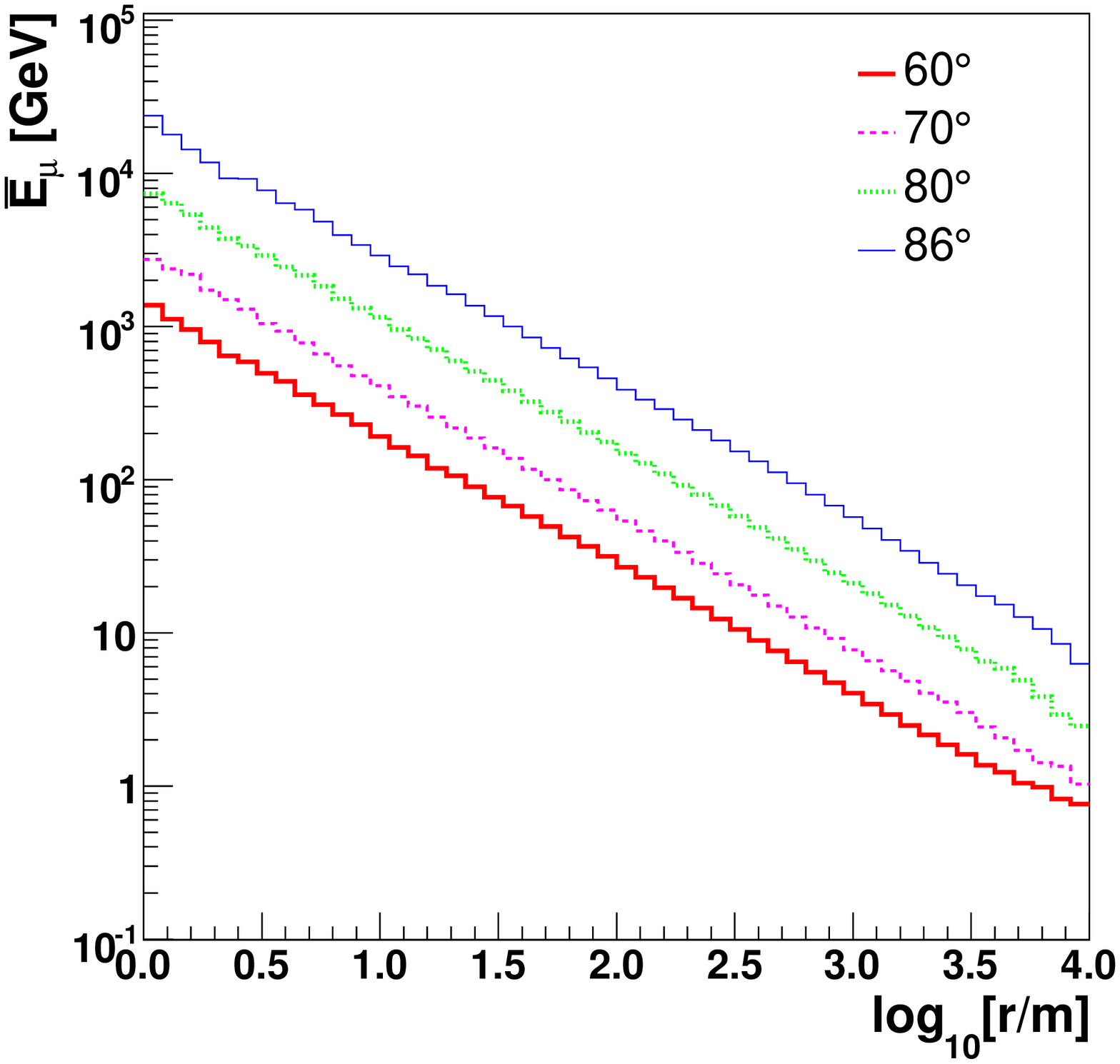}
\end{center}
\vspace{-15pt}
\caption{Left panel: Energy spectrum of muons at 30 m (open symbols)
  and 1000 m (full symbols) to the shower axis in the shower plane for
  showers at zenith angles 60$^{\circ}$ and 86$^{\circ}$. Right panel:
  Mean muon kinetic energy as a function of the distance to the shower
  axis in the shower plane for different zenith angles. The simulations
  were performed for 10 EeV proton showers with QGSJET01 hadronic
  model.}
\label{muonspectrum}
\end{figure*}

Since the ground arrays differ in detection thresholds and detector
response, we have performed particle energy cuts at ground level
corresponding to the particular case of water Cherenkov detector
thresholds: 264 KeV for electrons and positrons, 1.286 MeV for gammas
and 54.6 MeV for muons.  Note that the kinetic energy thresholds used
in the AIRES simulations for electromagnetic particles are all below 
the Cherenkov thresholds, implying that we are not artificially 
eliminating electrons, positrons or photons that could contribute
to the electromagnetic densities. Concerning muons, one can see 
by extrapolating the energy spectrum of muons shown in Fig.~\ref{muonspectrum} 
that the number of muons below 10 MeV in inclined showers 
-- not accounted for in our AIRES simulations but still being able
to produce electrons above the Cherenkov threshold in water --
represents a small fraction of the total number of muons.  

\subsubsection{Resampling procedure}

To calculate the particle densities induced by the electromagnetic and
muonic components of the shower, we perform a simple resampling
procedure (also called {\it unthinning} procedure).  The particle
density is calculated as the sum of the statistical weights assigned
by the thinning algorithm of the particles falling in a sampling
region divided by its area. The sampling area must be large enough so
that a significant fraction of particles falls inside it, but at the
same time it should be small enough so that particle properties are
representative of their expected properties in the particular region
of interest~\cite{Billoir:2008zz}. In this work, we have chosen two
different sampling regions in the plane perpendicular to the incoming
shower direction (shower plane). On the one hand following the
previous criterion,, we have considered particles falling in
concentric rings with width of 0.08 in $log_{10}r$ in the shower plane
to obtain the particle densities as a function of the distance to the
shower axis (lateral density). On the other hand, we have considered
square cells of $60 \times 60$ m$^{2}$ in the shower plane when
calculating the two dimensional distributions of the particle
densities.

\section{The electromagnetic and muonic components in inclined showers}

The conventional separation between vertical and inclined showers is
based on the zenith angle $\theta$ of the arrival direction of the
cosmic ray particle that induces the shower. This separation stems
from the different atmospheric grammage that the showers have to cross
before reaching the ground, which increases approximately as
$\sec{\theta}$. This fact implies that the showers arrive at ground at
different stages in their evolution depending on $\theta$.

A hadronic cosmic ray typically initiates an air shower at the top of
the atmosphere in the first few 100 g cm$^{-2}$. The EM component of
the shower rises as the shower develops and reaches a maximum at a
depth of $X_{\rm max}\sim \unit{800}\,{\rm g \,cm^{-2}}$ for a
\unit{10} {EeV} proton shower, close to the total vertical depth of
atmosphere for the altitude of the Pierre Auger Observatory.  After
$X_{\rm max}$, the EM component is rapidly absorbed in the atmosphere
mainly due to low-energy ionization processes and the photoelectric
effect. Meanwhile, non-decaying muons propagate almost unattenuated to
the ground, except for ionization energy losses and deflections in the
geomagnetic field. Consequently, a shower arriving with $\theta =
0^{\circ}$ reaches the ground shortly after shower maximum and the
electromagnetic component dominates at ground. However, for showers
arriving with $\theta>60^{\circ}$ the atmospheric slant depth
increases from $\sim\unit{1760}\,{\rm g \,cm^{-2}}$ to up to more than
$30000~{\rm g~cm^{-2}}$ for completely horizontal showers. This
results in the dominance of the muonic component at ground
\cite{Ave:2000dd} accompanied by a small electromagnetic component
consisting mainly of the remnant of the electromagnetic shower due to
cascading processes from $\pi^0$ decay of hadronic origin, and the
electromagnetic halo due to muon decay in flight and hard muon
interactions.

Low energy muons (below a few GeV) typically decay along their paths
to the ground generating small electromagnetic subshowers contributing
to the particle densities at ground.  This component mimics the muon
spatial distribution and is proportional to the muon density
\cite{AveThesis}.

Hard muon interactions (pair production, bremsstrahlung and hadronic
interactions) become more and more relevant as muon energy
increases. To illustrate when these processes are expected to
contribute larger to the EM halo, in the left panel of
Fig.~\ref{muonspectrum} we show the energy spectrum of muons at
distances of 30~m and 1000~m to the shower axis for 10 EeV
proton-induced showers at different zenith angles. As shown here,
these processes are expected to contribute to the EM halo especially
in highly inclined showers in which most of the muons are typically
very energetic (hundreds of GeV), because the lower energetic muons
typically decay before traveling the enlarged distances from their
production height to the ground. Hard muon interactions are also
expected to be more frequent close to the shower core where a larger
content of energetic muons is expected since energetic muons deviate
less from the shower axis.

\subsection{Asymmetry of the electromagnetic and muonic densities at ground level in inclined showers}

A vertical shower typically exhibits a symmetric pattern of particle
densities around the shower core in the shower plane.  In the case of
inclined showers there are several effects that produce an asymmetry
on the pattern of particle densities. The most important sources of
which are described below.

\subsubsection{Asymmetries due to the geomagnetic field}

\begin{figure*}
\begin{center}
\includegraphics[width=0.6\textwidth]{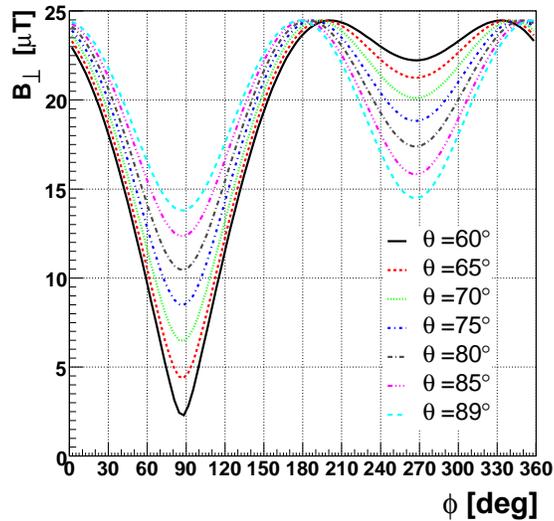}
\end{center}
\vspace{-15pt}
\caption{The component of the geomagnetic field perpendicular to the
  shower axis as a function of the shower azimuth angle for showers
  arriving with different zenith angles at the location of the Pierre
  Auger Observatory~\protect\footnotemark[2]. $\phi = 0^{\circ}$ corresponds to
  the geographical East, and $\phi = 90^{\circ}$ to the geographical
  North.}\label{compBperp}
\end{figure*}

Muons in inclined showers travel along sufficiently long paths in the
atmosphere to be affected by the Earth's magnetic field. Positive and
negative muons are deviated in opposite directions with respect to the
rectilinear trajectories they would follow in the absence of
geomagnetic field. 
\begin{figure*}
\begin{center}
\includegraphics[width=0.49\textwidth]{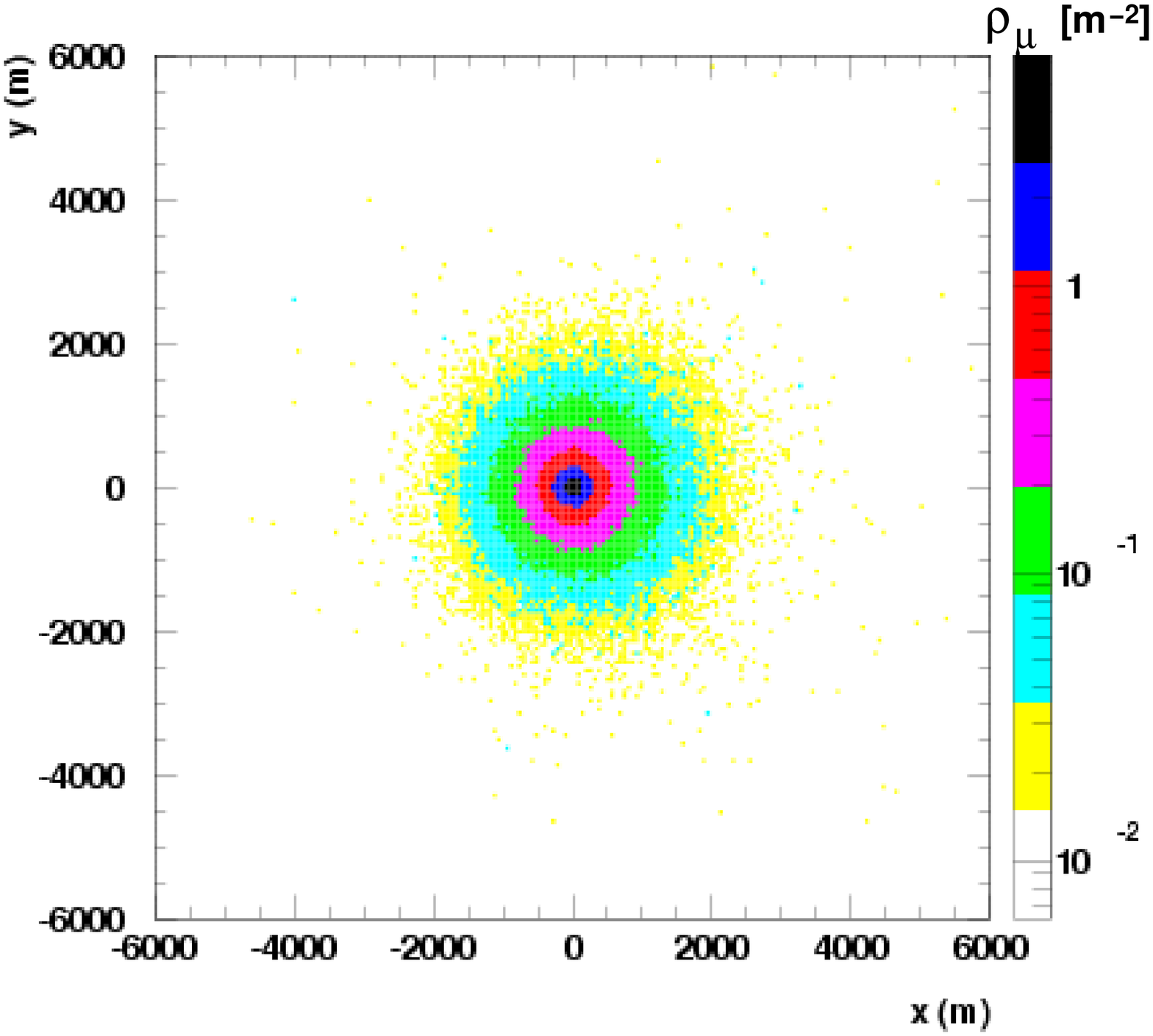}
\includegraphics[width=0.49\textwidth]{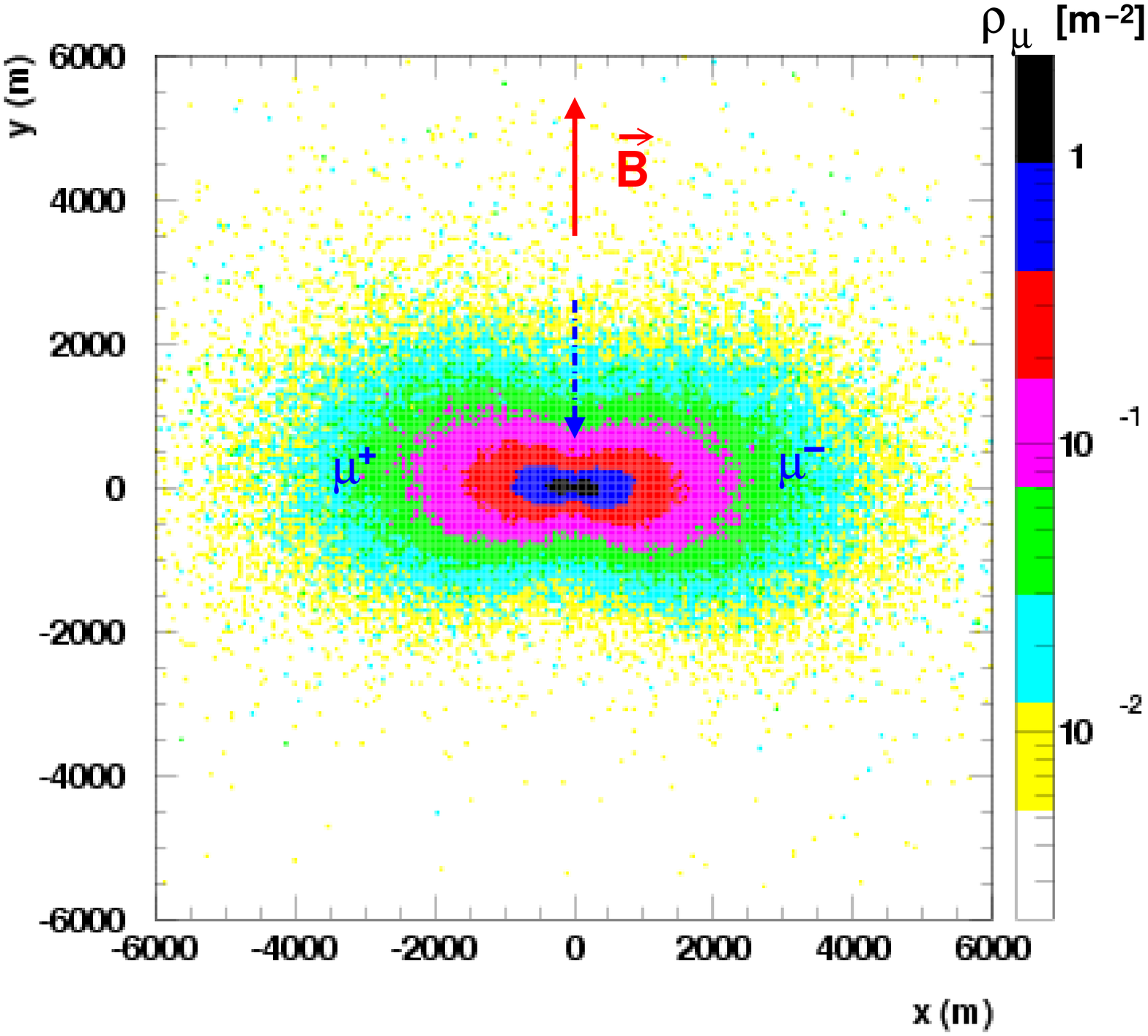}
\includegraphics[width=0.49\textwidth]{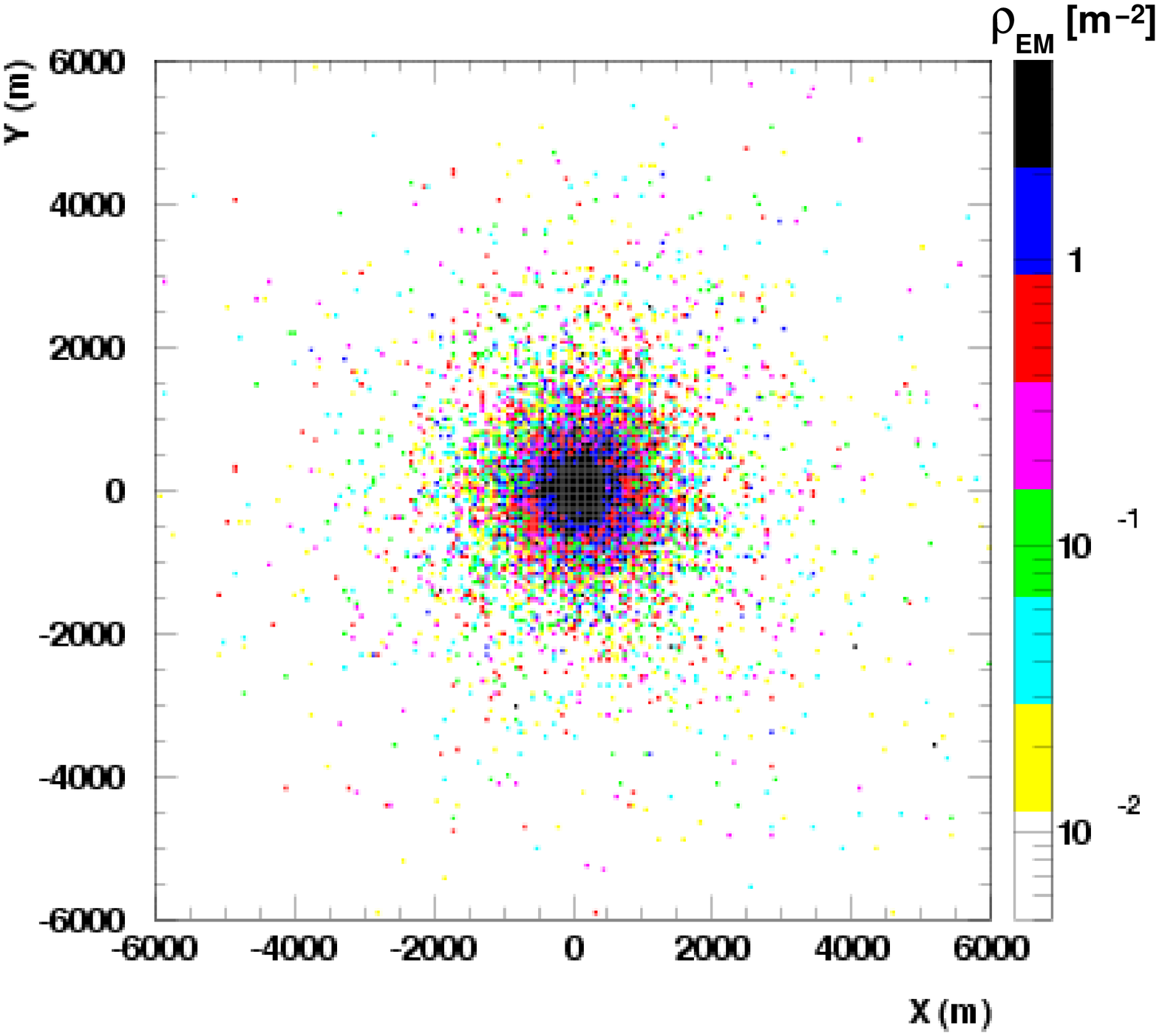}
\includegraphics[width=0.49\textwidth]{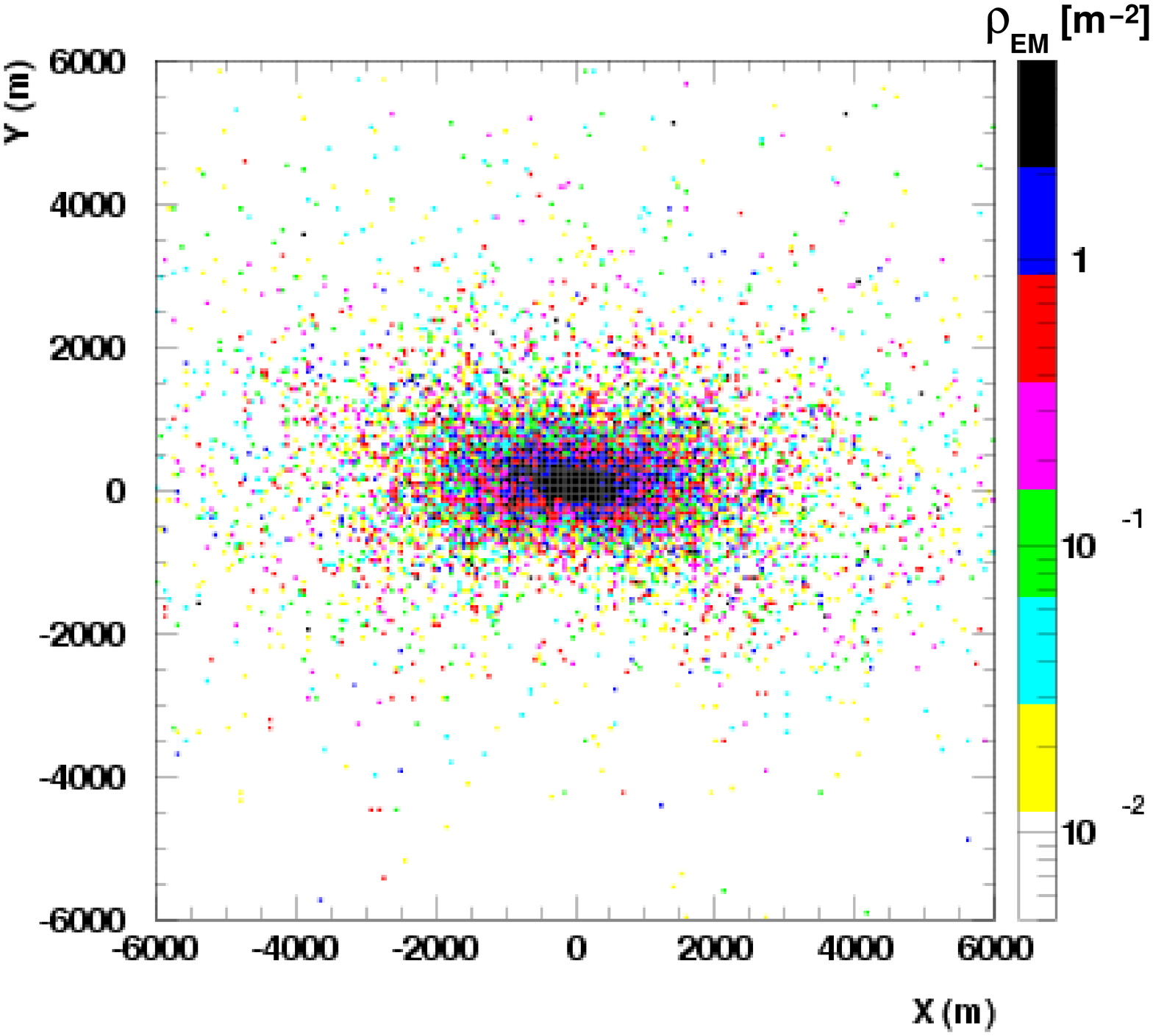}
\end{center}
\vspace{-10pt}
\caption{Maps of the muonic (top panels) and electromagnetic
    (bottom panels) densities in the shower plane in a $10$ EeV proton
    induced shower with a $\theta=86^{\circ}$ and $\phi=90^\circ$.
    Left panels: Densities without the effect of the geomagnetic
    field. Right panels: Maps with geomagnetic effect at the location
    of the Pierre Auger Observatory. The dashed arrow indicates the
    shower direction projected on the ground.  The solid arrow indicates
    the direction of the total magnetic field $\vec{\rm B}$ after
    projecting it on the ground. There is a component of $\vec{\rm B}$
    perpendicular to the shower axis. The simulations were performed
    for 10 EeV proton showers with the QGSJET01 hadronic model.\label{MuonMapWithandNoB}}
\end{figure*}
As a consequence the field distorts the patterns of
the muonic densities in the shower plane producing elliptical or even
2-lobed patterns in very inclined showers at ground, an effect
extensively studied in~\cite{Ave:2000xs,Hillas:1969zza}.  The degree
of distortion and the shape of the pattern depend on the strength and
orientation of the magnetic field with respect to the shower axis,
namely on the component of the field perpendicular to the shower axis
${\rm B}_{\perp}$, as well as on the distance traveled by the
muons. This introduces a dependence of the asymmetry on the shower
zenith ($\theta$) and azimuth ($\phi$) angles. As it was shown
in~\cite{Ave:2000xs} the effect on the muonic distributions is only
significant for $\theta \geq 75^{\circ}$.  In Fig.~\ref{compBperp} we
show ${\rm B}_{\perp}$ as a function of $\phi$ for the particular
location of the Southern Pierre Auger Observatory site.  For a given
zenith angle, the distortion is expected to be maximal close to $\phi
= 0^{\circ}$ and $\phi=180^{\circ}$ and minimal around $\phi =
90^{\circ}$. $\phi = 0^{\circ}$ points eastwards and is oriented
counterclockwise.

\footnotetext[2]{The geomagnetic field used in this work
    corresponds to the data of May 2006 extracted from the IGRF
    database~\cite{IGRF} included in AIRES. The strength of
    the field is 24.472 $\mu$T, the inclination angle below the
    horizon is 35.29$^{\circ}$ and the declination angle with respect
    to the geographical North is 2.91$^{\circ}$.}

As an example of the effect of the geomagnetic field on the particle
densities, in Fig.~\ref{MuonMapWithandNoB} we show 2-dimensional maps
of the muonic and electromagnetic particle densities in the shower
plane in a \unit{10}\,{EeV} proton induced shower arriving at $\theta
= 86^{\circ}$ and $\phi = 90^{\circ}$, with (right) and without (left)
the geomagnetic field induced.  The distortion of the cylindrical
symmetry of the muonic component is apparent, and since the
electromagnetic halo preserves the spatial distribution of muons in
inclined showers, the pattern of the electromagnetic particle density
exhibits a shape similar to the muonic one, although slightly blurred
due to multiple Coulomb scatterings suffered by the electrons before
reaching ground.

\begin{figure*}
\begin{center}
\includegraphics[width=0.6\textwidth]{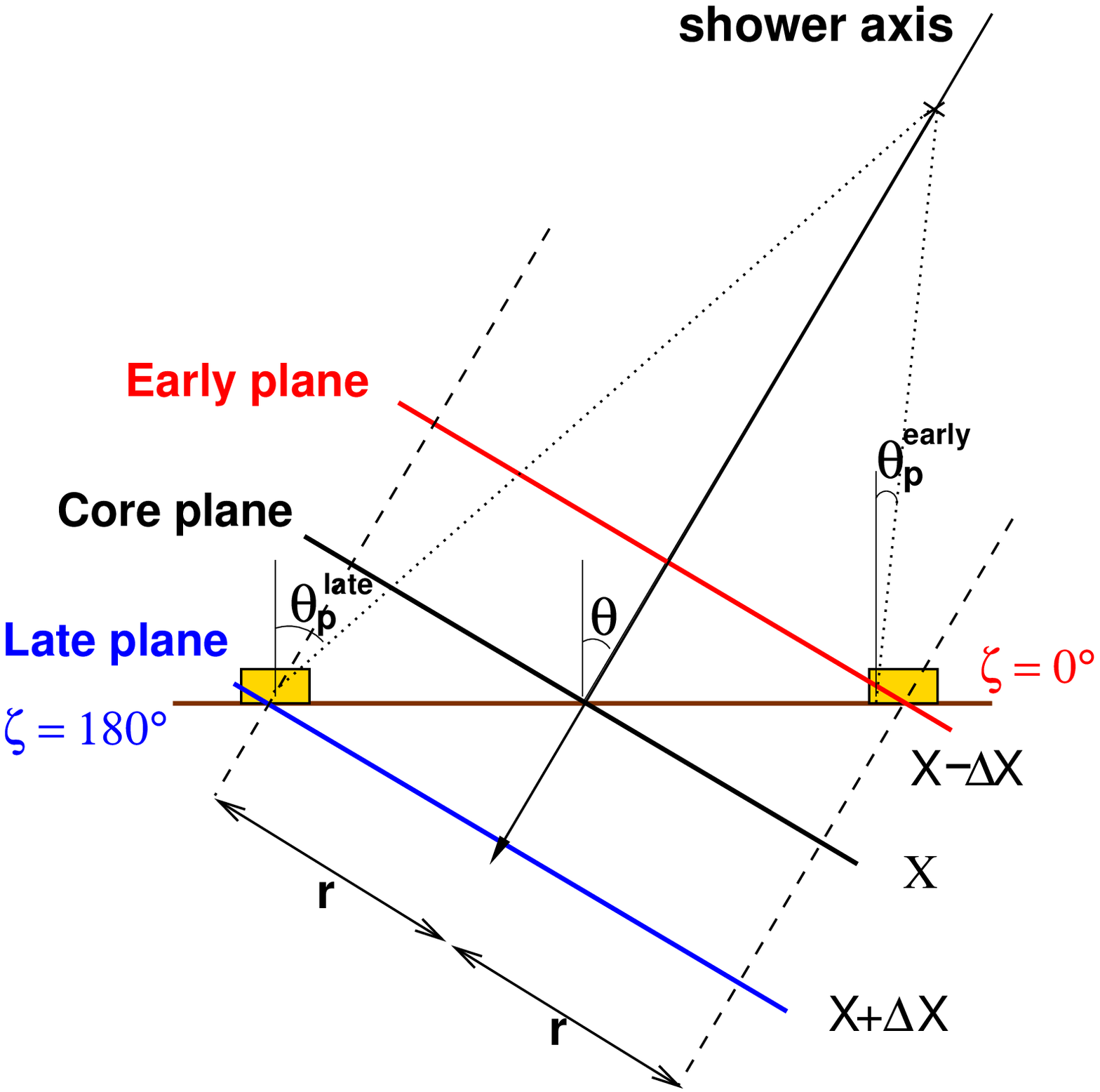}
\end{center}
\caption{Schematic view of an inclined shower reaching the
  ground. Three planes are displayed intersecting the ground plane,
  each one at different depth along the shower development: the early
  plane, the late plane and the shower core plane. The latter is also
  called shower plane.\label{AsymSchema}}
\end{figure*}

\subsubsection{Asymmetries due to geometrical and shower evolution effects}

Besides the asymmetries induced by the geomagnetic field, there is
also an azimuthal asymmetry in the muonic and electromagnetic particle
densities due to the combination of geometrical and shower evolution
effects~\cite{Dova:2001jy}. As illustrated in Fig.~\ref{AsymSchema}
shower particles do not travel parallel to the shower axis
in general and therefore they cross different amounts of atmospheric
depth before reaching ground. The crossed depth depends on the
azimuthal angle ($\zeta$) of the position on the ground at which the
particle arrives with respect to the incoming shower direction. In
particular, particles arrive at the ground in the \emph{early} region
of the shower (the portion of the shower front that hits the ground
first corresponding to $\zeta=0^\circ$) more vertically than those in
the \emph{late} region (corresponding to $\zeta=180^\circ$). This is
essentially the basis for the geometrical effect, which depends
strongly on the specific characteristics of the detectors sampling the
shower front. The asymmetry induced by the geometrical effect is
typically small in showers with high zenith angles. The more inclined
a shower, the more energetic the muons arriving at ground are and
hence the smaller the difference in the arrival angle distributions
between the early and late regions of the shower.

The asymmetry induced by the shower evolution can be understood as
follows. Particles at the same distance to the shower axis in the
shower plane, but arriving with different azimuthal angles $\zeta$
travel along different paths, and belong to different stages in the
evolution of the shower. The importance of this effect depends on the
evolution of the lateral particle distribution and on the attenuation
of the total number of particles with the atmospheric depth. In
Fig.~\ref{AsymSchema} the detector in the early region of the shower
is hit by a younger stage in the evolution of the shower than the
detector in the late region. More quantitatively, for instance in a 10
EeV proton shower at $\theta = 60^{\circ}$, there is a difference of
$2\,\Delta X \sim 370$ g cm$^{-2}$ (for notation see
Fig.~\ref{AsymSchema}) between the atmosphere crossed by particles
produced at the same height, traveling in straight line and hitting
the ground in the late and early regions at a distance $r = 1000$ m to
the core in the shower plane. This difference increases with the
distance from the shower core.

\begin{figure*}
\begin{center}
\includegraphics[width=\textwidth]{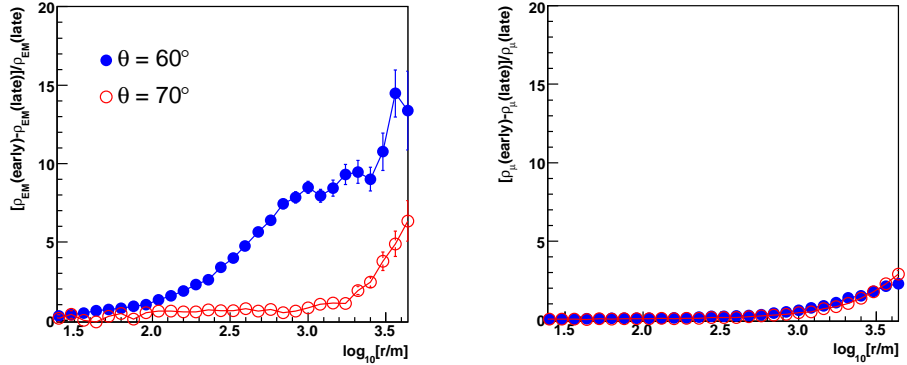}
\end{center}
\vspace{-15pt}
\caption{ Left panel: Early-late asymmetry of the
    electromagnetic particle density as a function of the distance
    from the core in the shower plane for showers at $\theta =
    60^{\circ}$ (full circles) and $\theta = 70^{\circ}$ (empty
    circles). Right panel: Early-late asymmetry of the muon particle
    density as a function of the distance from the core in the shower
    plane for showers at $\theta = 60^{\circ}$ (full circles) and
    $\theta = 70^{\circ}$ (empty circles). The simulation was
    performed for 10 EeV proton showers with the QGSJET01 hadronic
    model.\label{AsymDensit}}
\end{figure*}

The asymmetry induced by the evolution of the shower affects more the
electromagnetic component produced by $\pi^{0}$ decay than the muonic
component or its associated electromagnetic halo. The reason is that
this component is exponentially suppressed after the shower maximum,
and small variations of the depth crossed by the shower induce large
differences in the number of electromagnetic particles on the
ground. However, the muonic component is less attenuated and therefore
the asymmetry induced by this effect is smaller. As a consequence,
shower evolution is expected to induce a much smaller asymmetry in the
density in showers with zenith angles $\theta > 70^{\circ}$, because
the electromagnetic component from $\pi^{0}$ decay is practically
suppressed, and the electromagnetic halo is simply following the
pattern of the muonic component where the asymmetry is small. This is
shown in Fig.~\ref{AsymDensit} where we plot the relative difference
between the electromagnetic and muonic particle densities in the early
($\zeta \approx 0^{\circ}\pm15^\circ$) and late ($\zeta =
180^{\circ}\pm 15^\circ$) regions in the shower plane as obtained in
our simulations. For the moment the geomagnetic field effect is
neglected in this study. In the left panel of Fig.~\ref{AsymDensit} it
can be seen that for showers at $\theta = 60^{\circ}$ the difference
in the electromagnetic density between the early and the late regions
is very large even at small distances to the core (for example a factor 1.2 at $r
\sim$ 100 m), because the electromagnetic component from $\pi^{0}$
decay is still significant in the early region, and it is
significantly absorbed before reaching ground in the late region. This
difference increases steeply with distance $r$. However, at $\theta =
70^{\circ}$ the difference between the densities in the early and late
regions is smaller, because the electromagnetic component from
$\pi^{0}$ decay is absorbed at all $\zeta$ and only the
electromagnetic halo remains, except for at very far distances from the
core where, besides inheriting the already quite large asymmetry due to the 
muonic component, there is still some remnant of the electromagnetic shower 
present in the early region but not in the late one. This explains why the early-late
asymmetry follows essentially the behaviour of the corresponding
muonic density (right panel of Fig.~\ref{AsymDensit}). In the muonic
case, the relative difference between the early and late densities is
always small regardless of zenith angle.

\subsection{The lateral distribution of the electromagnetic and muonic components in inclined showers}

The electromagnetic and muonic particle densities have a
characteristic behaviour with distance to the shower axis, shower
zenith angle and azimuthal position with respect to the incoming
shower direction. Also the different contributions to the
electromagnetic particle densities differ from each other as will be
shown below.

Firstly, we have studied the EM and muon number densities as a
function of the distance to the shower axis (lateral distributions) in
10 EeV proton-induced showers at different zenith angles, averaging
over the azimuthal angle $\zeta$ on the ground. We have also
parameterised the lateral distributions of the particle densities as a
function of shower zenith angle for practical applications, using 10
EeV proton showers simulated with the hadronic model QGSJET01 as
reference. These parameterisations are presented in Appendix A. The
results from these parameterisations are shown as solid lines in the top
panels of Fig.~\ref{EmMuSignal} compared to the simulations. The simulations
are reproduced by the fits without significant deviations as shown in the
bottom panels of Fig.~\ref{EmMuSignal}  ($< 10\%$ for muons and $< 30\%$ for
the EM component).

\subsubsection{Muonic component}

\begin{figure*}
\begin{center}
\hspace{-10pt}
\includegraphics[width=0.49\textwidth]{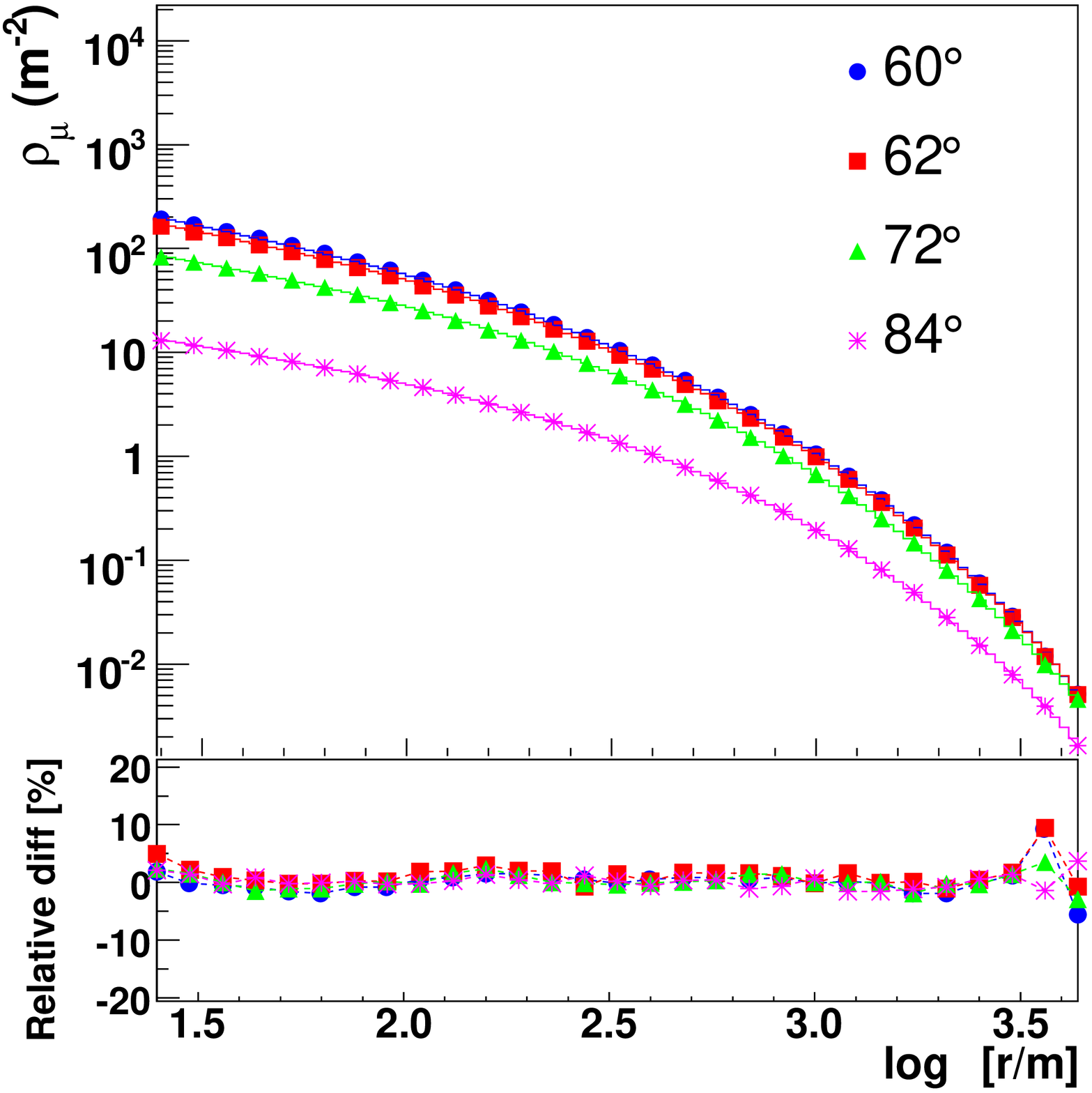}
\includegraphics[width=0.49\textwidth]{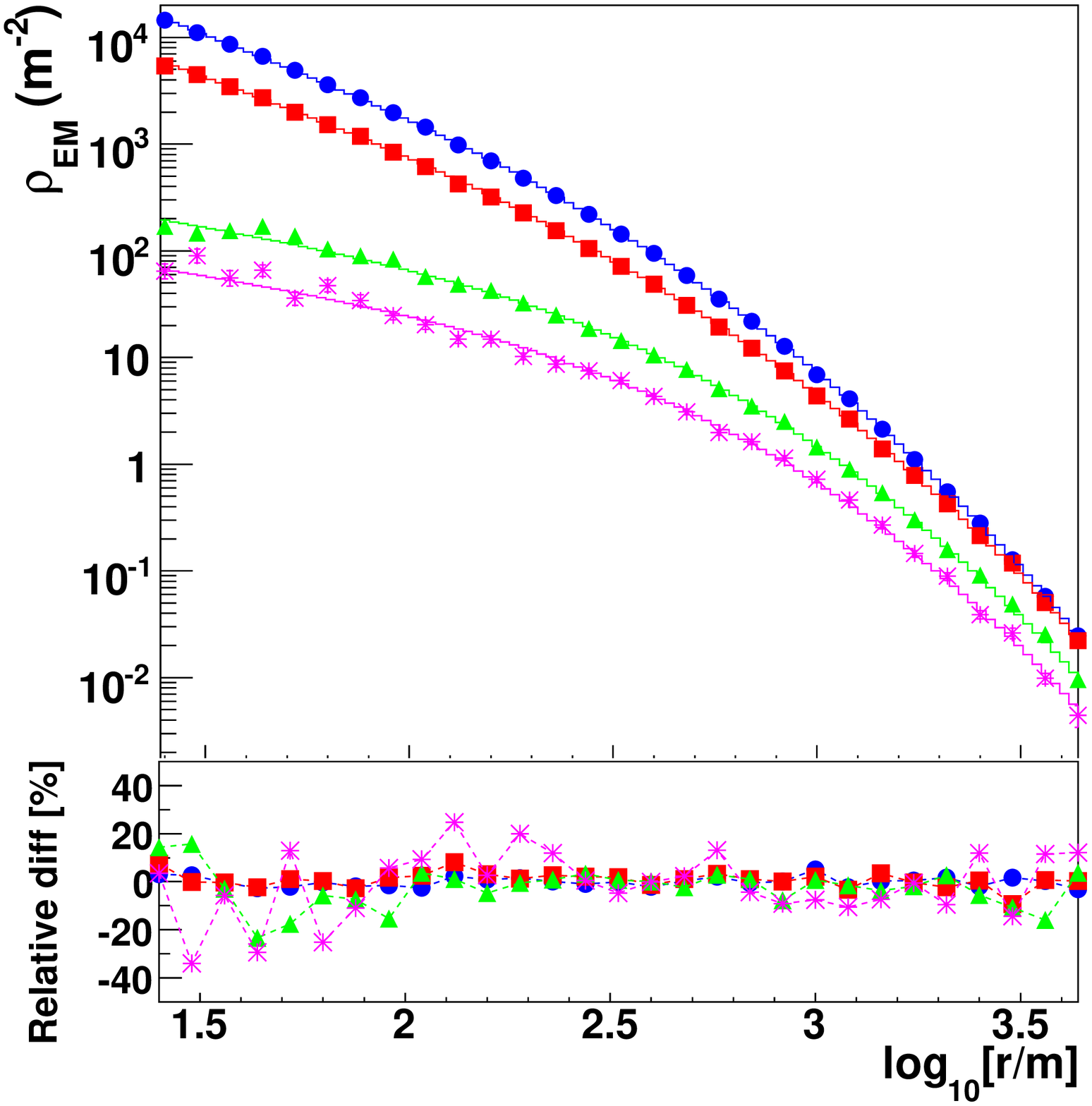}
\end{center}
\vspace{-15pt}
\caption{Top panels: Muonic (left panel) and electromagnetic (right panel)
    particle densities as a function of the distance to the shower axis in the
    shower plane for 10 EeV proton showers at different zenith angles
    simulated with the QGSJET01 hadronic model. For a fixed $r$ the densities
    were averaged over the azimuth angle $\zeta$.  The solid lines indicate
    the results of the fitting functions in Appendix A. Bottom panels: Relative
    difference in $\%$ between the muonic (left panel) and electromagnetic
    (right panel) densities as obtained with the parameterisations in Appendix
    A with respect to the simulations.\label{EmMuSignal}}
\end{figure*}

As shown in the left panel of Fig.~\ref{EmMuSignal}, the muonic
density $\rho_\mu$ decreases with $\theta$, because muons need to
travel larger distances before reaching ground arriving at larger
distances to the core and being spread over a larger area. This is mainly due
to the transverse momentum inherited from its parent hadron. Muons of
the lowest energies typically decay, decreasing further the density
with theta, especially at large distances. 

Both effects also explain the behaviour of the muon energy
distribution with $\theta$ and distance to the shower axis
shown in Fig.~\ref{muonspectrum}. For example, in the right
panel of this figure one can see that the mean muon energy increases
with $\theta$ for a fixed $r$, because only the more
energetic muons survive and besides, these deviate more from the
shower axis.

\begin{figure*}[h]
\begin{center}
\hspace{-20pt}
\includegraphics[width=0.6\textwidth]{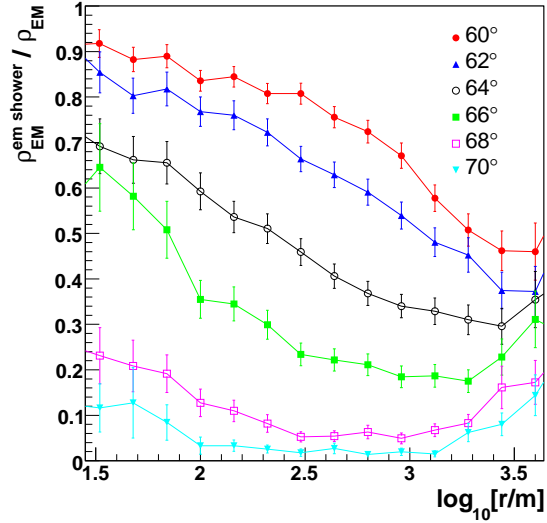}
\end{center}
\vspace{-15pt}
\caption{Contribution of the electromagnetic particle density due to
  $\pi^0$ decay of hadronic origin in the shower (remnant EM shower)
  to the total EM density as a function of distance to shower axis for
  different zenith angles. This component was obtained in special
  simulations in which muons are not explicitly followed so that they
  cannot contribute to the EM halo (see text for
  details).\label{EMRemnant}}
\end{figure*}

\subsubsection{Electromagnetic component}

The behaviour of the electromagnetic lateral distribution in inclined
showers (right panel of Fig.~\ref{EmMuSignal}), can be qualitatively
understood as a combination of the different behaviour of the two
contributions to the EM component namely, that produced by $\pi^0$
decay (from hadronic origin) and the electromagnetic halo (from muonic
origin). 

Separating the two main contributions to the total EM density
is not possible in AIRES simulations because the information on the
mother particle (a $\pi^0$ or a muon) producing the EM subshower is
lost. However we have devised a procedure to obtain in an approximate
way the various contributions to the total EM density. The total EM
density $\rho_{\rm EM}$ can be obtained as:
 
\begin{equation}
\rho_{\rm EM} = \rho_{\rm EM}^{\rm em~shower} + 
\rho_{\rm EM}^{\rm \mu~decay} + 
\rho_{\rm EM}^{\rm \mu~int}
\end{equation} 
where $\rho_{\rm EM}^{\rm em~shower}$ is the EM density due to decay of
$\pi^0$s produced in hadron and meson interactions, $\rho_{\rm EM}^{\rm
  \mu~decay}$ is the EM density due to muon decay and $\rho_{\rm EM}^{\rm
  \mu~int}$ is the EM density due to muon bremsstrahlung, pair production and
nuclear interactions. Using AIRES we have performed a special subset of shower
simulations in which we artificially set the muon energy threshold above which
muons are explicitly followed in the simulations to a very high energy (10
TeV), and at the same time we set the muon lifetime to infinity so that muons
never decay. The last condition needs to be forced otherwise muons are
artificially decayed in AIRES. The EM density obtained in this way has no
contribution from muons (i.e. $\rho_{\rm EM}^{\rm \mu~decay} + \rho_{\rm
  EM}^{\rm \mu~int}\simeq 0$) and can only be generated in $\pi^0$ decays
(produced in hadron and meson interactions). In Fig.~\ref{EMRemnant} we show
the lateral distribution of the EM density as obtained in these simulations
for different zenith angles. It can be seen that as the zenith angle increases
from $60^\circ$ the EM density due to $\pi^0$ decay is increasingly absorbed,
until the zenith angle reaches $\sim70^\circ$ and it practically disappears,
except for a small contribution still reaching ground very near the core at
$r<100$ m and far from it at $r>2$ km reaching only the early region of the
shower.

For $\theta > 70^{\circ}$ the EM component from $\pi^{0}$ decay is
negligible and the EM halo dominates at essentially all distances to
the core. This is reflected in the fact that the EM lateral
distribution follows the behaviour of the muonic one. This is more apparent
in Fig.~\ref{EmMuRatio} where the ratio of the electromagnetic and
muonic densities is shown. Note that the results shown in this figure
are obtained with standard AIRES simulations.

\section{The ratio of the electromagnetic to muonic contributions to the densities}

We have also studied the behaviour of the ratio of the electromagnetic to the muonic densities:
 
\begin{equation}
\ R_{\rm EM/\mu} = \rho_{\rm EM} /\rho_{\mu}
\label{ratioEMMU}
\end{equation}

\begin{figure*}
\begin{center}
\hspace{-20pt}
\includegraphics[width=0.51\textwidth]{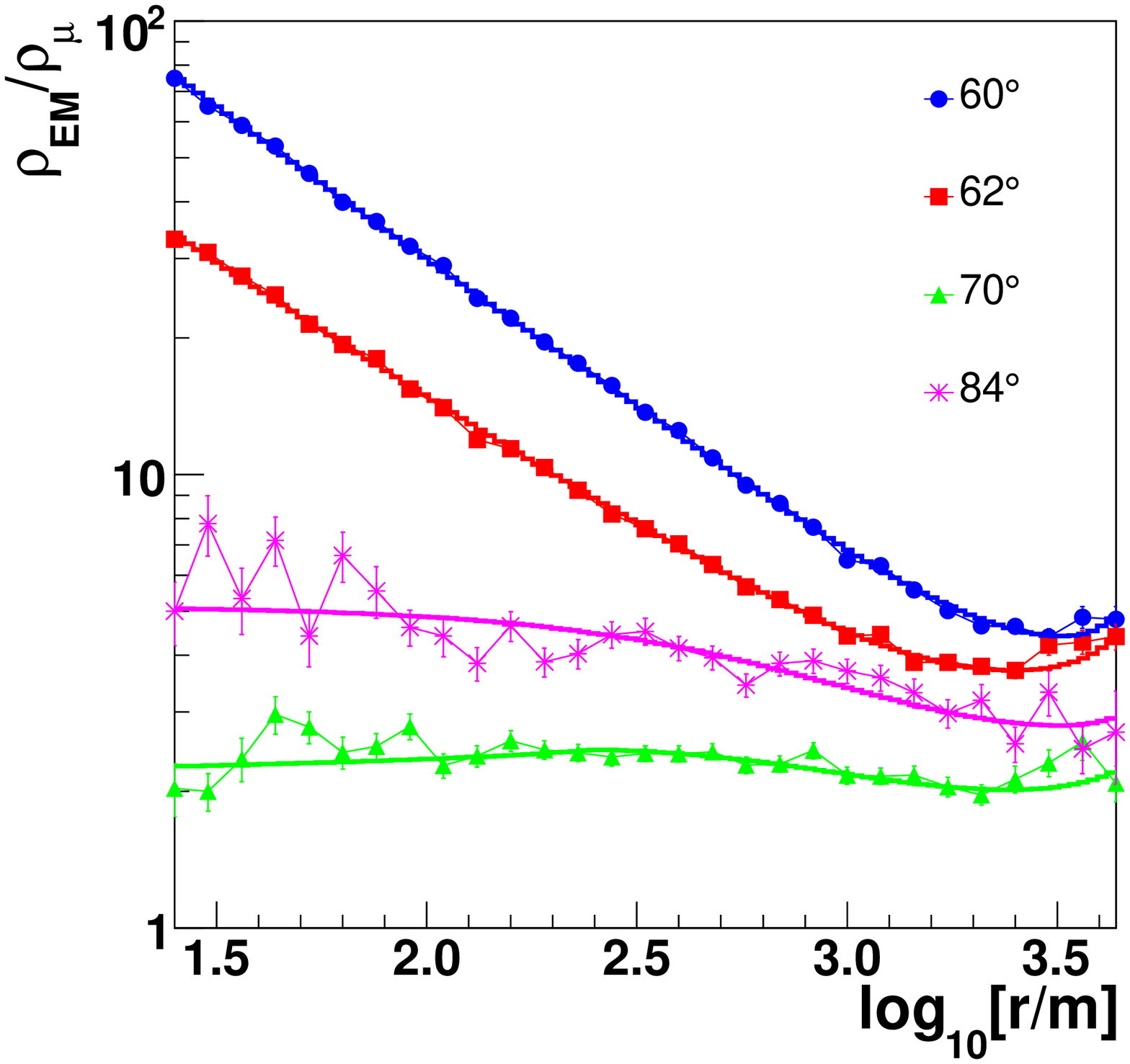}
\includegraphics[width=0.51\textwidth]{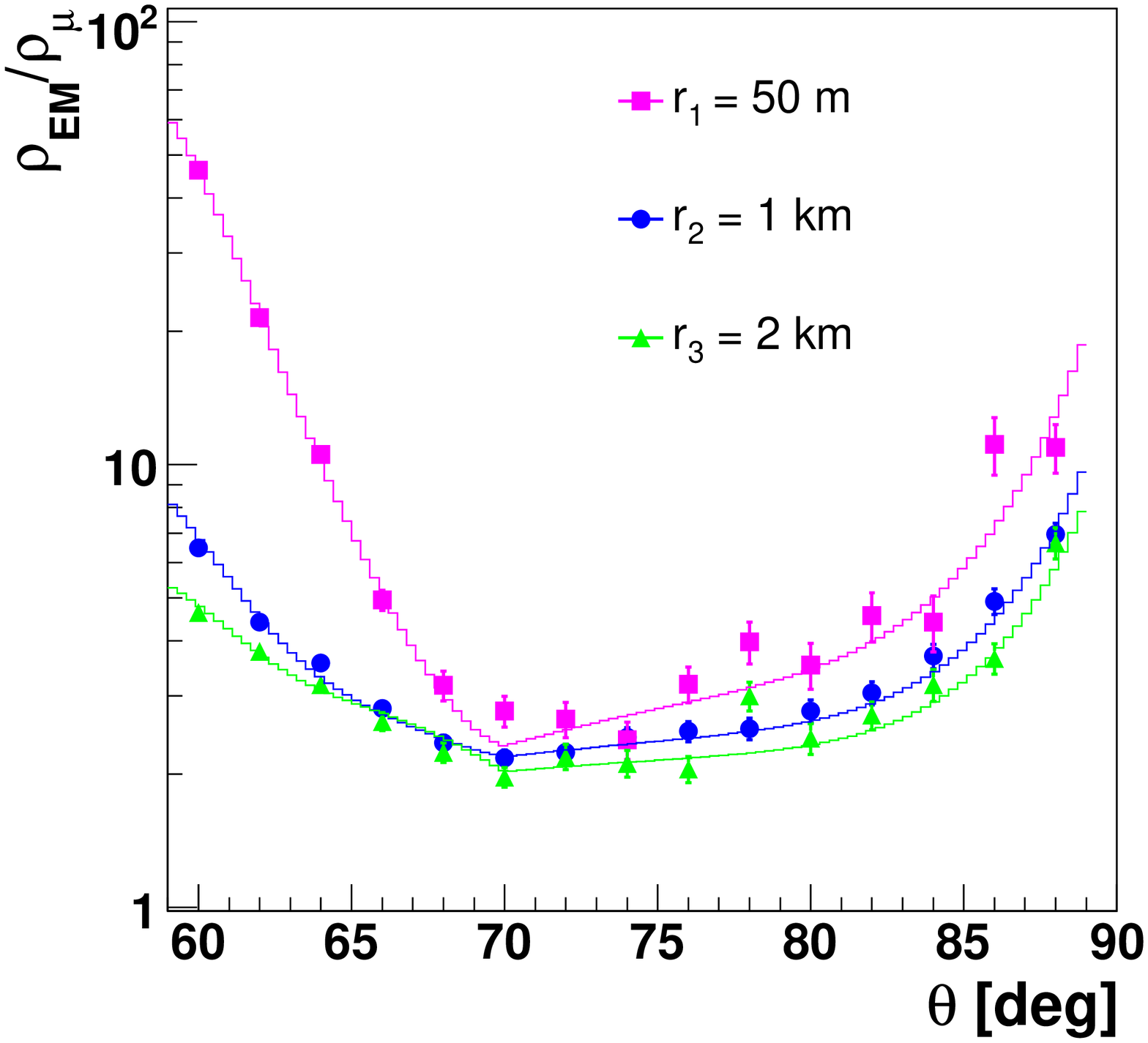}
\end{center}
\vspace{-10pt}
\caption{Left panel: The ratio of the electromagnetic to muonic
  particle densities as a function of the distance from the shower
  axis in the shower plane for different zenith angles. Right panel:
  The ratio of the electromagnetic to muonic particle densities as a
  function of the shower zenith angle for different distances from the
  shower axis. The simulations were performed for 10 EeV proton
  showers in absence of geomagnetic field. The solid line indicates
  the ratio of the results given by the fitting functions in Appendix
  A.\label{EmMuRatio}}
\end{figure*}

In Fig.~\ref{EmMuRatio}, we show $R_{\rm EM/\mu}$ averaged over
azimuth angle $\zeta$ as a function of the distance to the core $r$
for different shower zenith angles $\theta$ and as a function of the
$\theta$ for different fixed distances as obtained in the simulations,
and the ratio predicted by the parameterisations proposed in Appendix
A. Near the core, the ratio decreases with zenith angle from $\theta =
60^{\circ}$ to $\sim 70^{\circ}$ due to the increasing absorption of
the EM component from $\pi^0$ decay, and then increases again with
$\theta$ as can be seen in Fig.~\ref{EmMuRatio}, mainly due to muon
hard interactions processes that are expected to dominate near the
core in very inclined showers. Far from the core the lateral
distribution of the ratio tends to flatten due to the dominant
contribution of the EM halo produced by muon decay in flight. The
larger the zenith angle, the closer to shower core the ratio levels
off. The slight increase of the ratio for $\theta \lesssim 68^{\circ}$
and far from the core ($r \gtrsim 2$ km ) is attributed to the
combination of two effects, one is that the number of low energy muons
decreases more rapidly at large distances because they decay before
reaching the ground, and only energetic muons survive, and on the
other hand the presence of the contribution to the EM density due to
$\pi^0$ decay, particularly in the early region of the shower.

\subsection{Azimuthal asymmetry in the ratio of the electromagnetic to muonic densities}

In addition to the dependence on zenith angle and distance to shower
core, we have also studied the azimuthal asymmetry of $R_{\rm
  EM/\mu}$. For the moment the effect of the geomagnetic field is
neglected.

\begin{figure*}
\begin{center}
\includegraphics[width=0.49\textwidth]{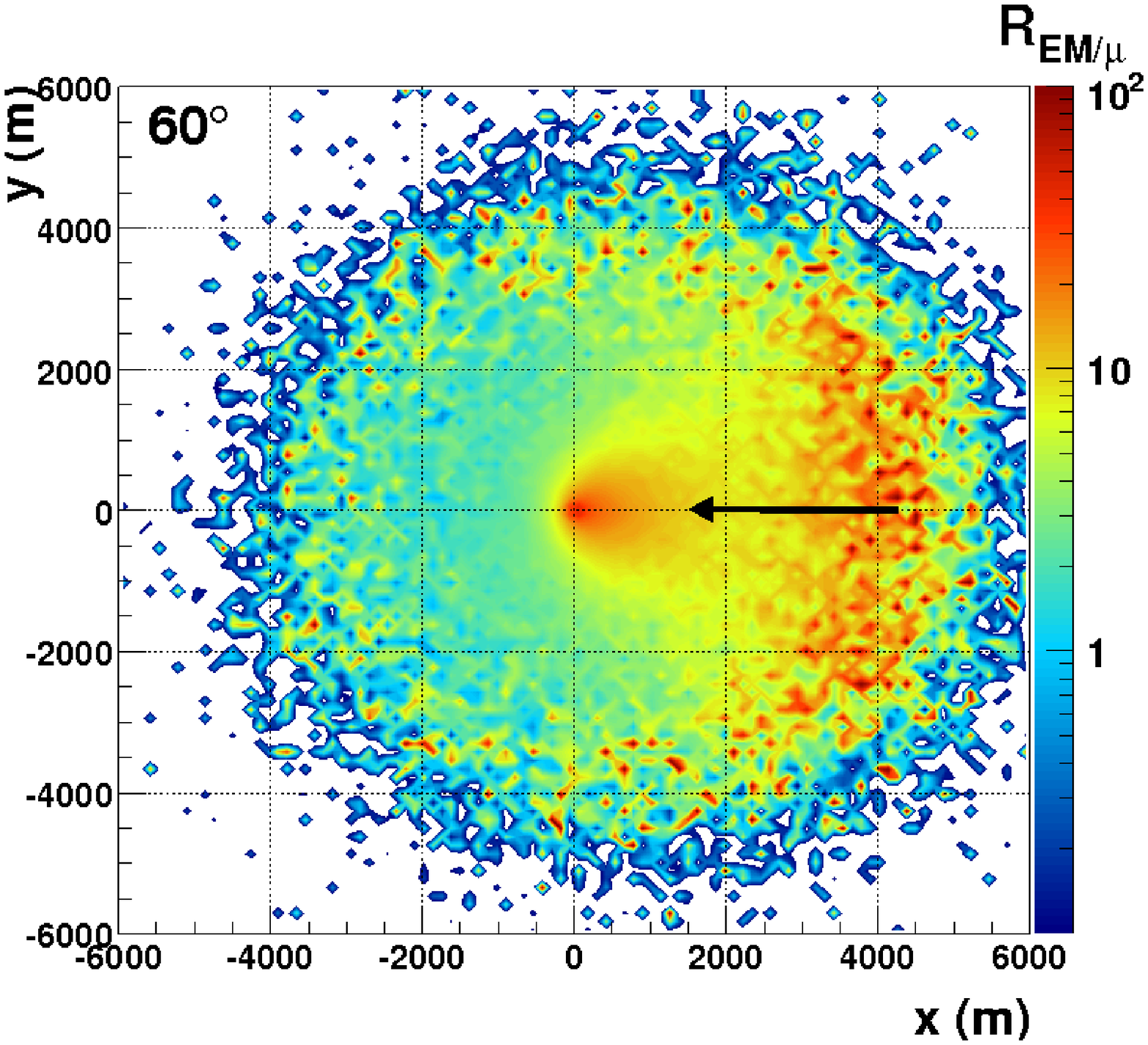}
\includegraphics[width=0.475\textwidth]{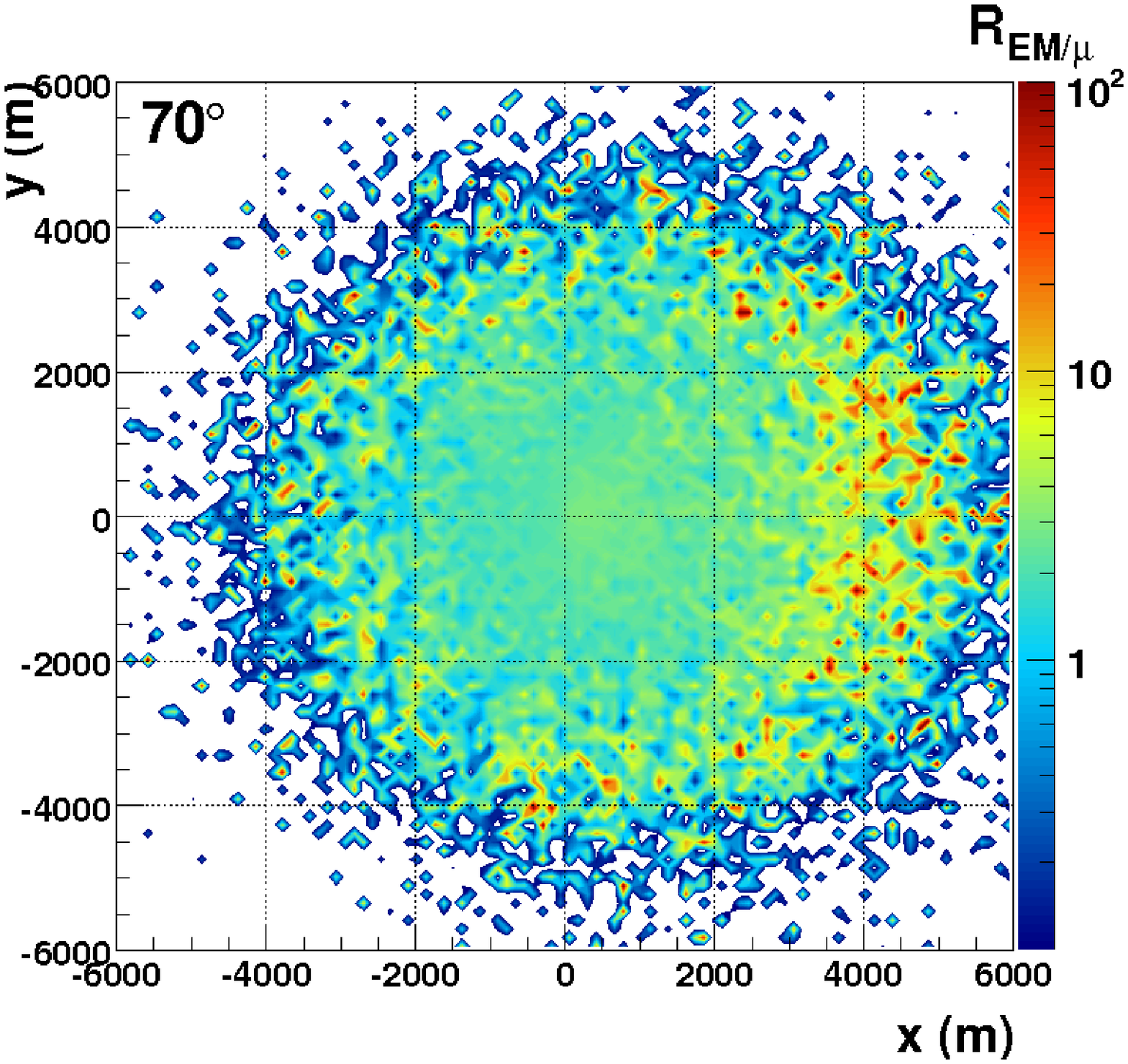}
\end{center}
\caption{Map of the ratio $R_{\rm EM/\mu}$ in the shower plane for 10
  EeV proton showers at $\theta = 60^{\circ}$ (left panel) and $\theta
  = 70^{\circ}$ (right panel). The arrow indicates the shower
  direction projected on the ground. The shower core hits ground at
  the position (x=0,y=0).\label{RatioMap}}
\end{figure*}

In the left panel of Fig.~\ref{RatioMap}, we show the 2-dimensional
map of the ratio $R_{\rm EM/\mu}$ in the shower plane for 10 EeV
proton showers at $\theta = 60^{\circ}$. The shower incoming direction
is from East to West in the figure. As expected, there is a clear
azimuthal asymmetry at a fixed distance to the core. The contribution
of the electromagnetic component is larger in the early
region. However, as shown in the right panel of Fig.~\ref{RatioMap},
at zenith angles $\theta \geq 70^{\circ}$ the azimuthal asymmetry is
less significant because only muons accompanied by the electromagnetic
halo arrive at ground. Since these two components approximately have
the same asymmetry (section 3.1.2 ), the final asymmetry is
practically canceled out when making their ratio.

To study further the azimuthal dependence of the asymmetry we divide
the shower plane in bins of width $\Delta\zeta = 30^{\circ}$ centered
around $\zeta$, and we calculate the lateral distributions of the
ratio in each bin for a fixed zenith angle: $R_{\rm
  EM/\mu}(r,\theta,\zeta)$, and we compare these distributions to the
distribution $\langle{R}_{\rm EM/\mu}\rangle(r,\theta)$ obtained
averaging over $\zeta$. For this purpose we define the asymmetry
parameter $\Delta_{\zeta}$ as
 
\begin{equation}
\ R_{\rm EM/\mu}(r,\theta,\zeta) = \langle{R}_{\rm EM/\mu}\rangle (r,\theta) \times (1 + \Delta_{\zeta})
\label{AsymratioEMMU}
\end{equation}

\begin{figure*}
\begin{center}
\hspace{-15pt}
\includegraphics[width=0.49\textwidth]{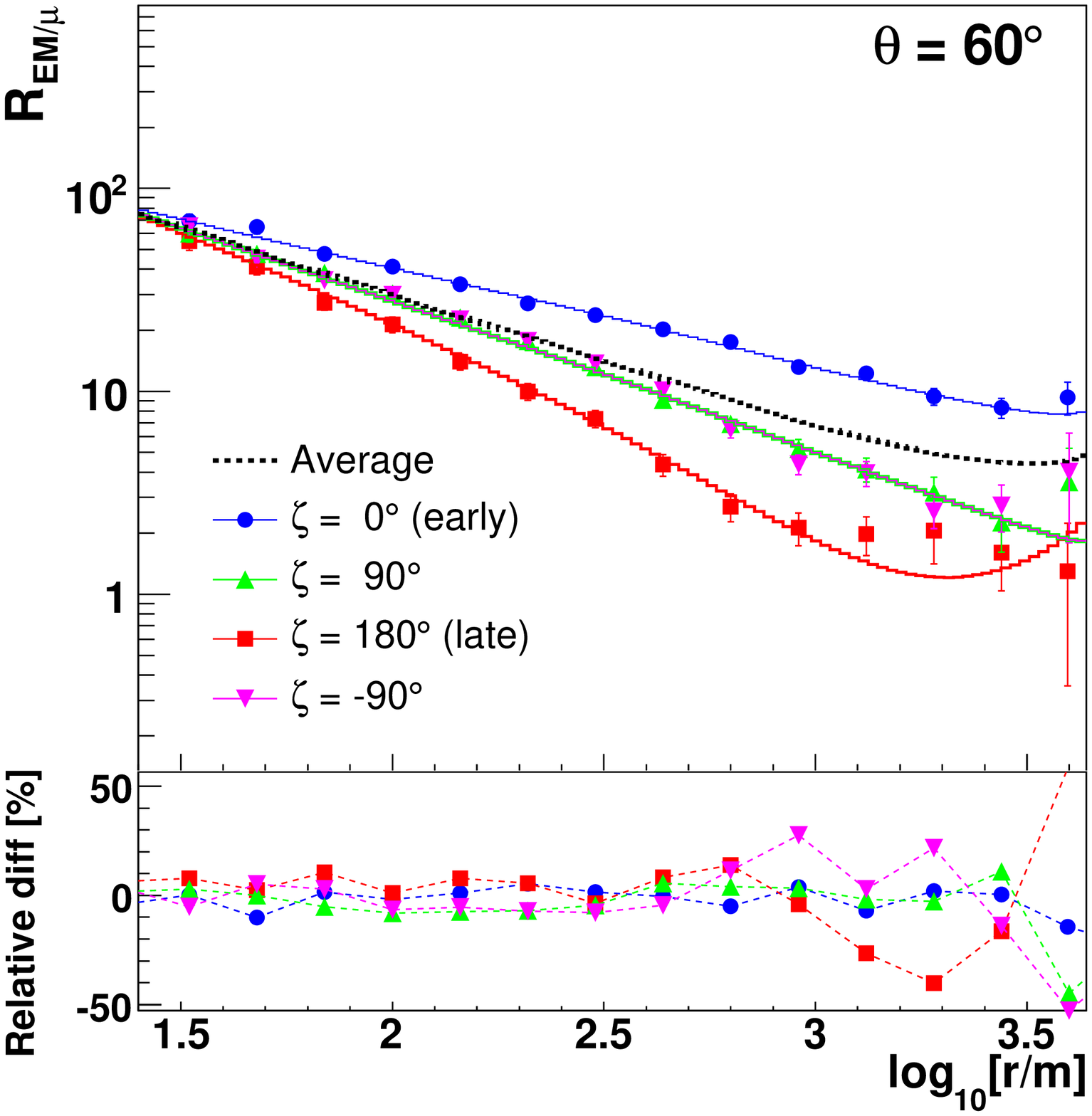}
\includegraphics[width=0.5\textwidth]{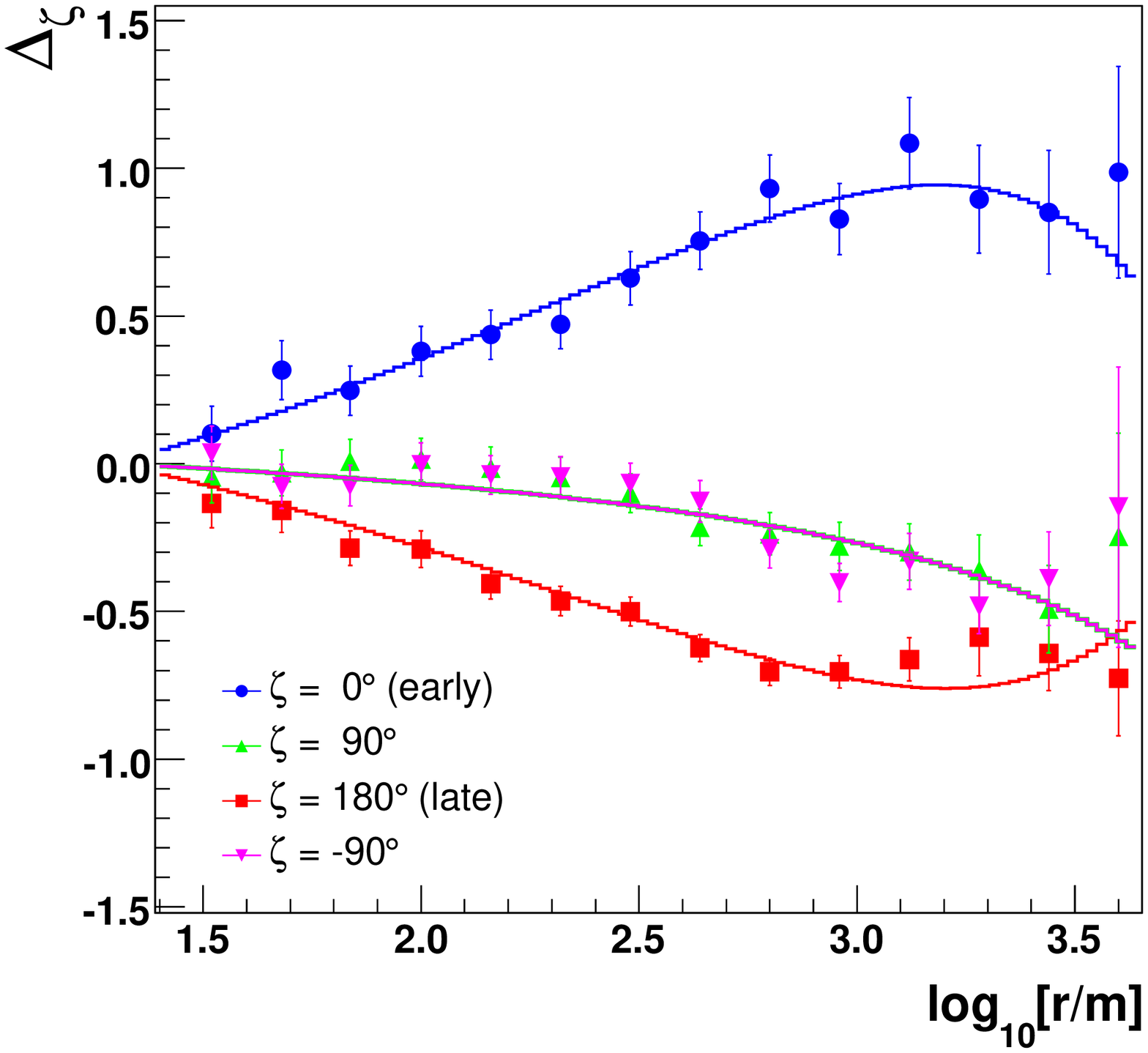}
\end{center}
\caption{Left panel: The ratio $R_{\rm EM/\mu}$ as a function of the distance
  from the shower axis in the shower plane in different bins in $\zeta$ for 10
  EeV proton showers with $\theta = 60^{\circ}$. The relative difference in
  $\%$ between the ratio as obtained with the parameterisation in Appendix B
  with respect to the simulations is shown in the bottom panel. Right
  panel: Asymmetry of the lateral distribution of the ratio $R_{\rm EM/\mu}$
  in different $\zeta$ bins. The size of the bins is $\Delta\zeta=30^\circ$
  centered at $\zeta$. The solid line indicates the ratio of the results given
  by the fitting functions in Appendix B.\label{ExampleAsyEmMuRatio}}
\end{figure*}

In Fig.~\ref{ExampleAsyEmMuRatio}, we show the lateral distribution of
$R_{\rm EM/\mu}$ in different $\zeta$ bins compared to the mean value
(left panel) and their corresponding asymmetry parameter
$\Delta_{\zeta}$ (right panel) for showers at $\theta = 60^{\circ}$.
$\vert\Delta_{\zeta}\vert$ increases with distance to the core and it
is larger in the early region than in the late region as expected.

\begin{figure*}
\begin{center}
\includegraphics[width=0.7\textwidth]{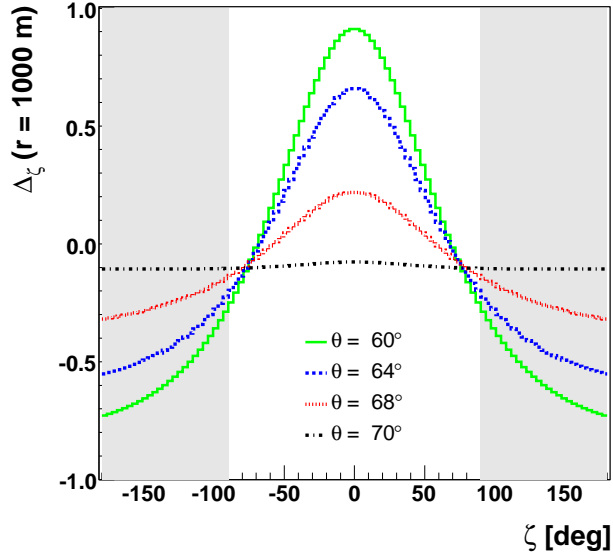}
\end{center}
\vspace{-15pt}
\caption{Asymmetry of the ratio $R_{\rm EM/\mu}$ as a function
    of the azimuth angle in the shower plane $\zeta$ at $r = 1000$ m
    and for different $\theta$. The shaded area indicates the late
    region of the shower and the remaining area corresponds to the
    early region.\label{AmplitudeAsy}}
\end{figure*}

In addition to the dependence on position in the shower plane
($r,\zeta$), the asymmetry parameter $\Delta_{\zeta}$ also depends on
the shower zenith angle. In Fig.~\ref{AmplitudeAsy} we show
$\Delta_{\zeta}$ as a function of $\zeta$ for a fixed distance $r =
1000$ m and different zenith angles. The amplitude of the asymmetry
decreases as the zenith angle increases for the reasons explained
above. This plot illustrates the importance of accounting for the
asymmetry in the ratio when dealing with inclined showers with
$60^{\circ} < \theta < 70^{\circ}$.

We have parameterised the asymmetry $\Delta_{\zeta}$ as a function of the
distance to the core, zenith angle and the azimuthal angle $\zeta$ accounting
for all the dependences above (see Appendix B). As an example, we show the
results of the fit as solid lines in Fig.~\ref{ExampleAsyEmMuRatio}. The 
simulations are reproduced by the fit without significant deviations as shown in the
bottom panel of Fig.~\ref{ExampleAsyEmMuRatio}. In the
parameterisation, we have assumed that the azimuthal asymmetry 
$\Delta_{\zeta}\simeq 0$ is negligible when $\theta > 68^{\circ}$.

\subsection{Geomagnetic field effect on the ratio of the electromagnetic to muonic densities}

As discussed before (section 3.1.1), the muon distributions are
distorted by the presence of the geomagnetic field for zenith angles
greater than $\theta\sim 75^{\circ}$.

\begin{figure*}
\begin{center}
\includegraphics[width=0.49\textwidth]{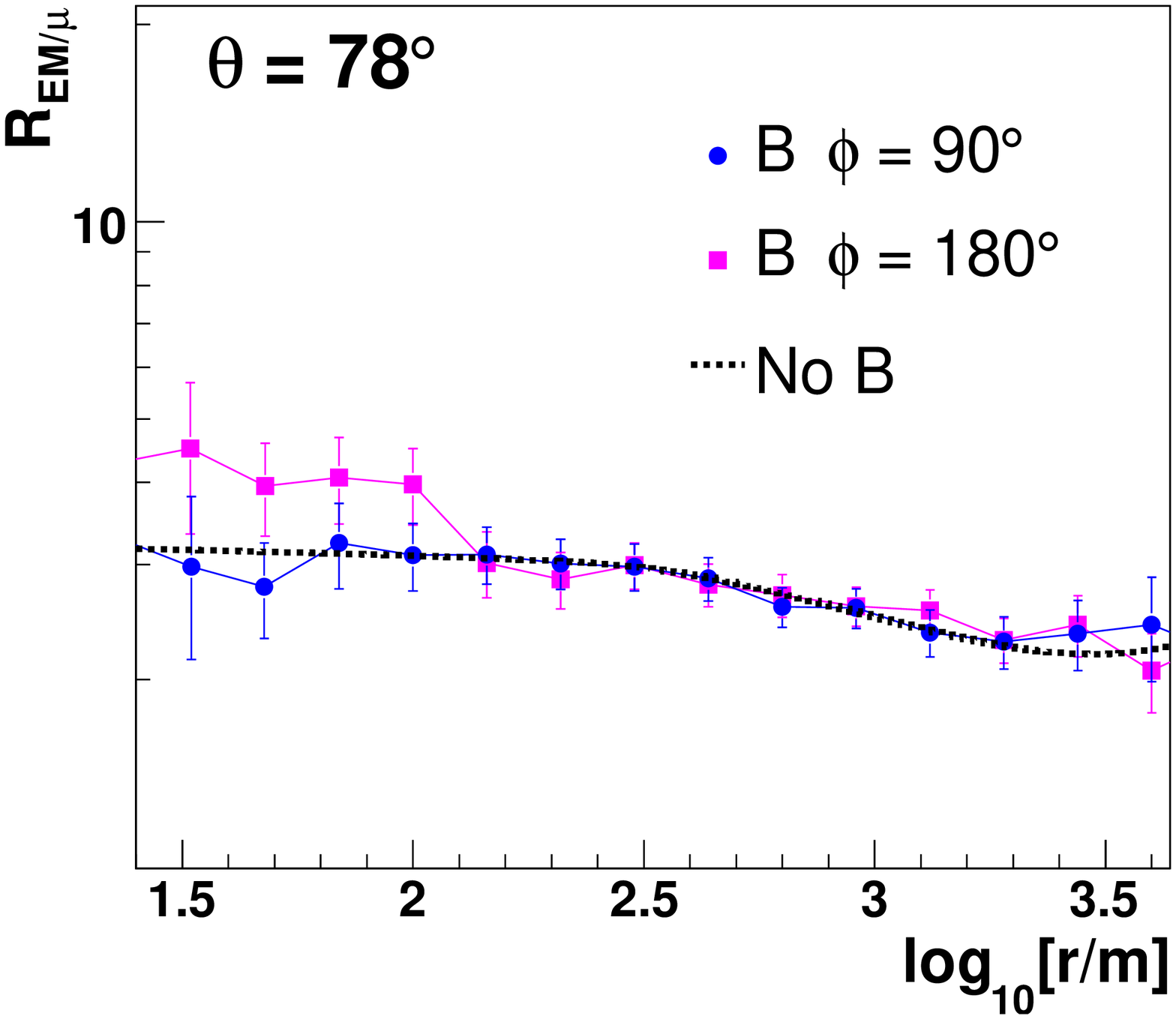}
\includegraphics[width=0.49\textwidth]{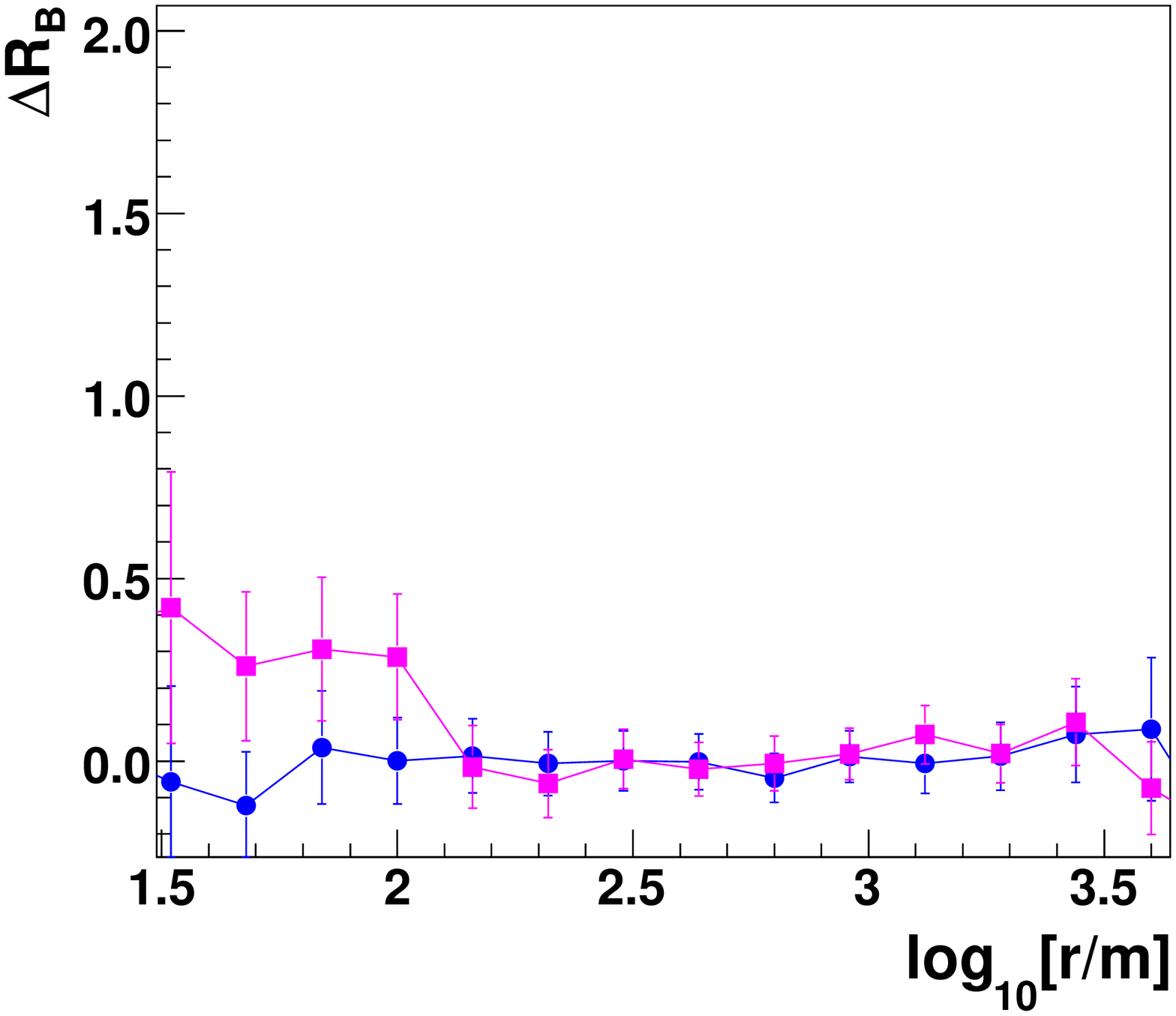}
\includegraphics[width=0.49\textwidth]{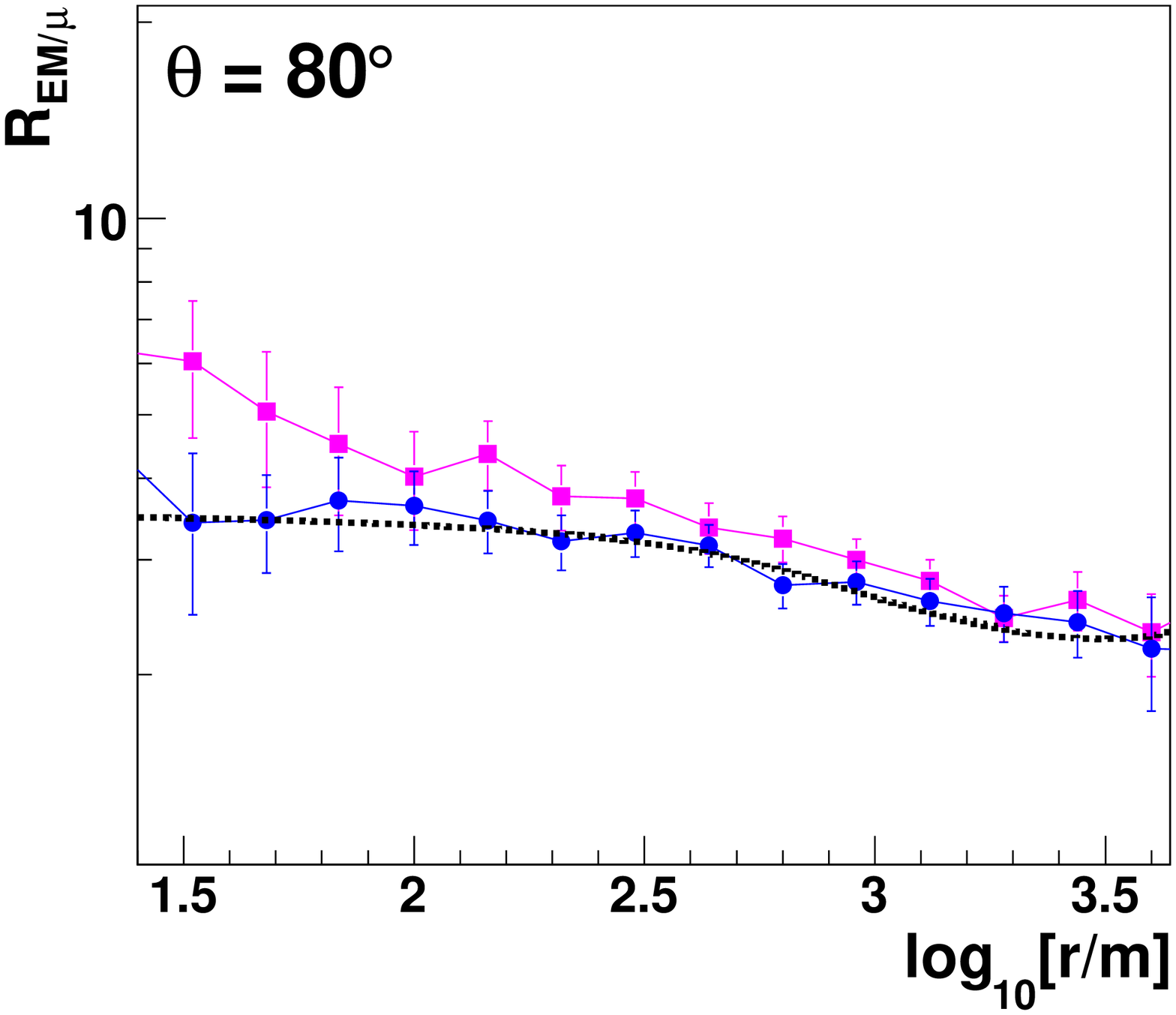}
\includegraphics[width=0.49\textwidth]{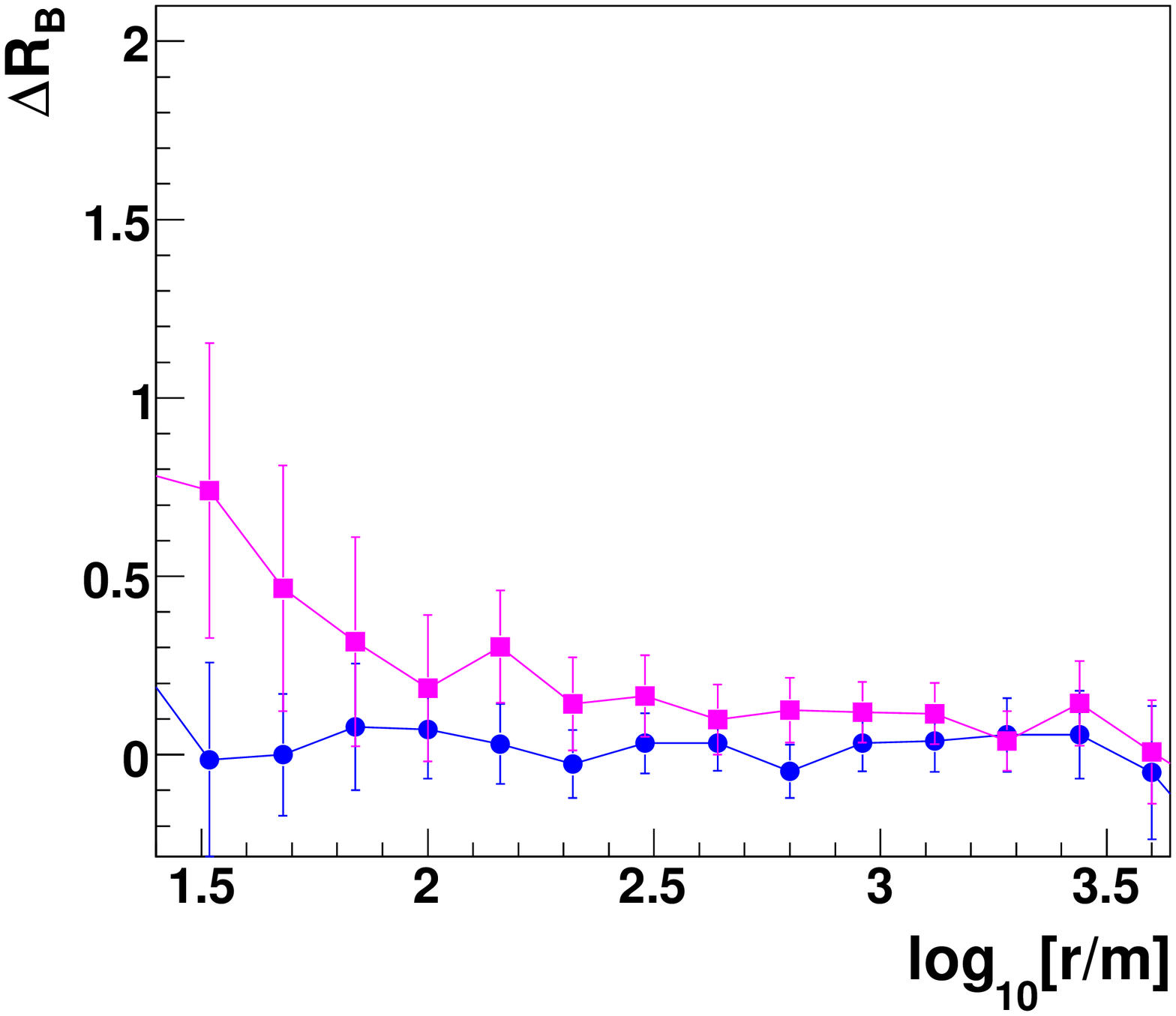}
\includegraphics[width=0.49\textwidth]{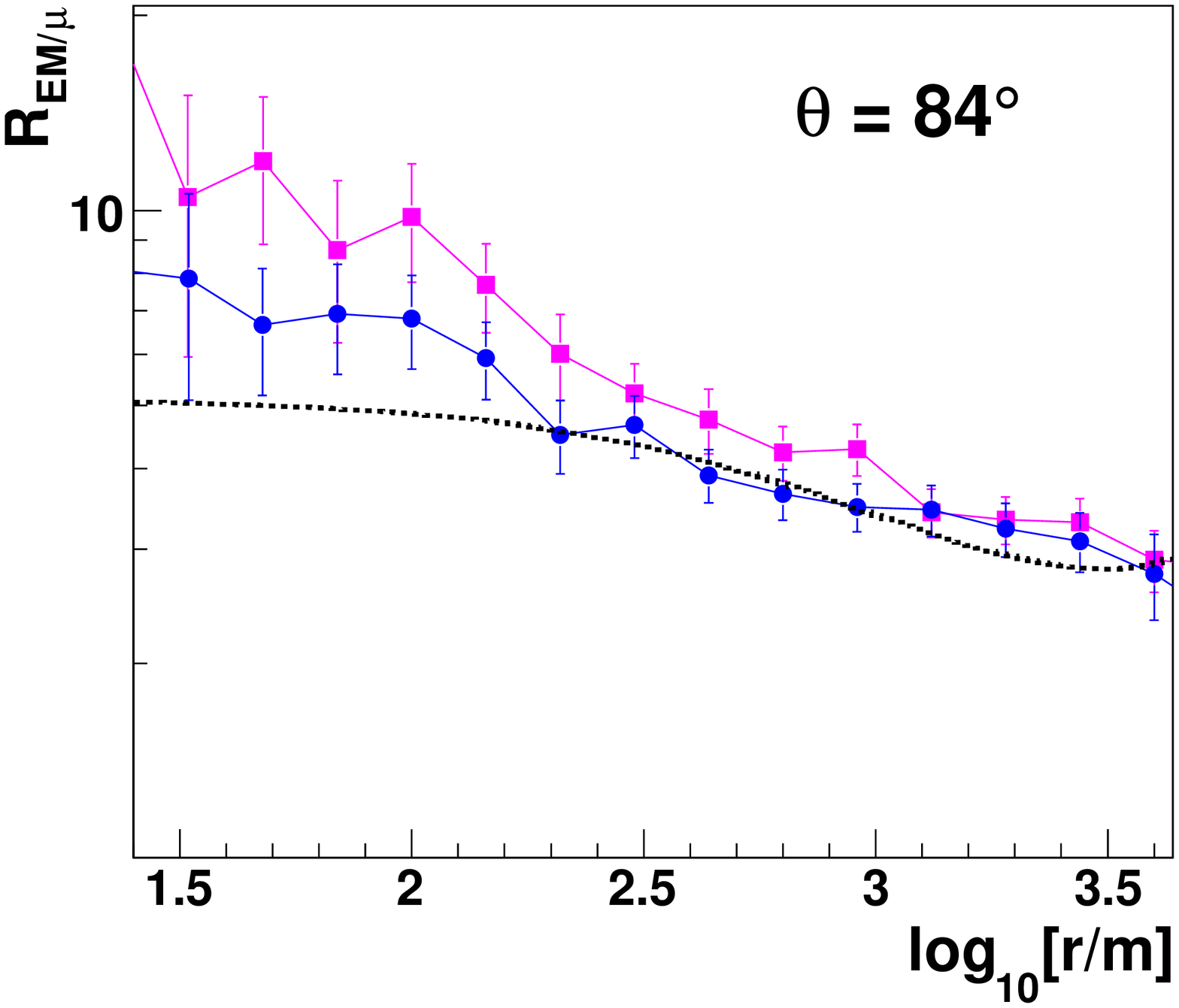}
\includegraphics[width=0.49\textwidth]{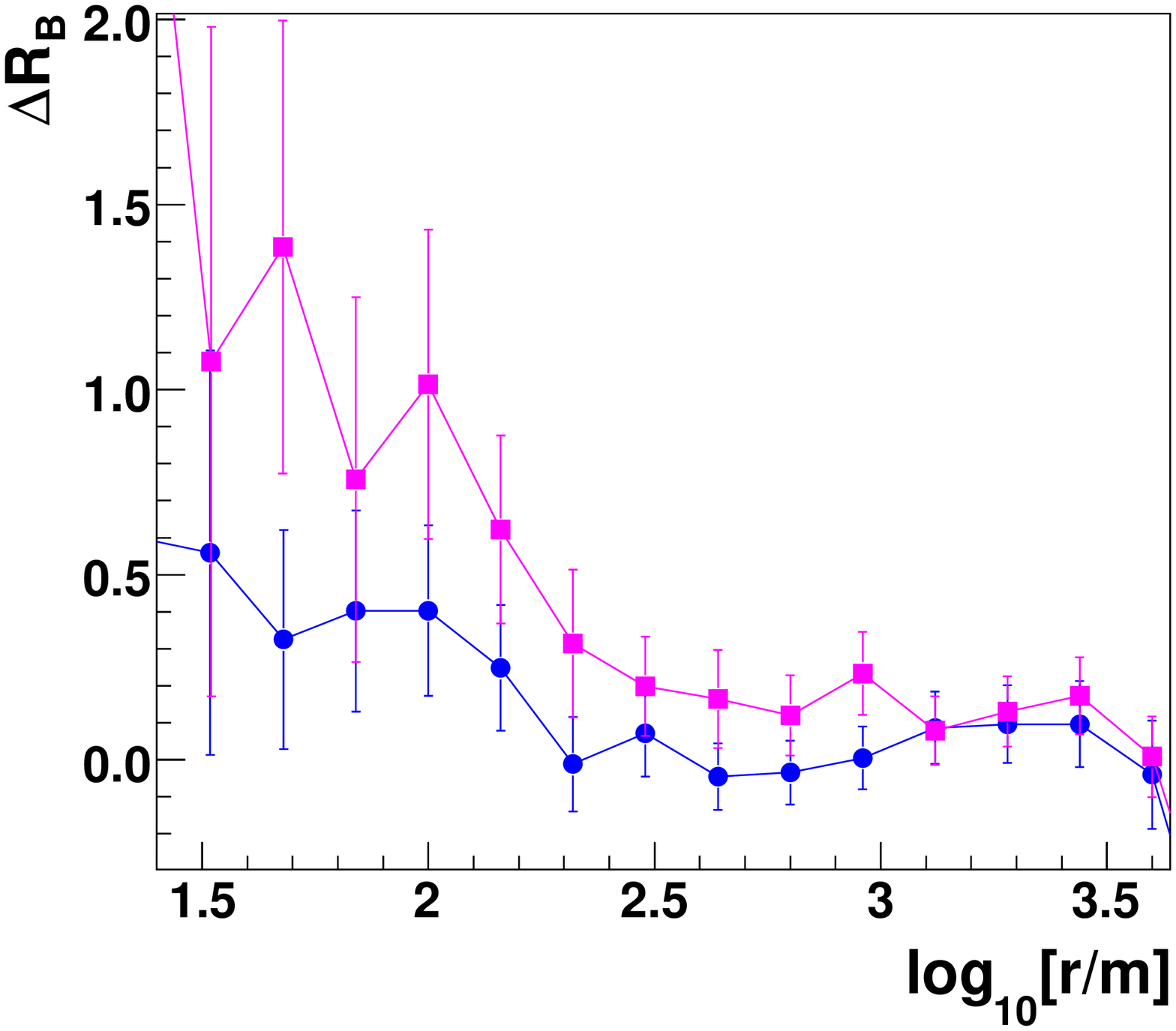}
\end{center}
\caption{Left panel: The ratio $R_{\rm EM/\mu}$ as a function of
    the distance from the shower axis in the shower plane for 10 EeV
    proton showers for different $\theta$ and $\phi = 90^{\circ}$
    (circles) and $\phi = 180^{\circ}$ (squares) in the presence of
    the geomagnetic field. The lateral behaviour of the ratio without
    the effect of the geomagnetic field is also shown (dashed line).
    Right panel: The relative difference with respect to the ratio as
    obtained without the geomagnetic field (see text).\label{BeffRatio}}
\end{figure*}
At these angles, the dominant
contribution to the electromagnetic component at ground is due to the
electromagnetic halo, which inherits the muon spatial distribution and
is proportional to the muonic density, and we expect the ratio of the
electromagnetic to muonic densities to maintain the symmetry in the
azimuthal angle $\zeta$. For this reason, we have only studied the
effect of the geomagnetic field on the ratio averaging over the
azimuthal angle $\zeta$ in the shower plane, for different shower
zenith $\theta$ and azimuth $\phi$ angles. More precisely, for each $\theta$ we
study the extreme cases where the effect is minimal, $\phi \approx
90^{\circ}$, and maximal, $\phi \approx 0^{\circ}$ (or $180^{\circ}$),
where $\phi$ is defined as in Fig.~\ref{compBperp}. In the left
panels of Fig.~\ref{BeffRatio} we show the lateral behaviour of
$R_{\rm EM/\mu}$ in the presence of the geomagnetic field for 10 EeV
proton showers. We also plot the relative difference between $R_{\rm
  EM/\mu}$ with and without geomagnetic field effect, namely:
 
\begin{equation} 
\Delta R_{B} = \frac{R_{\rm EM/\mu}(\phi)- \langle{R}_{\rm EM/\mu}\rangle(B
   = 0)}{\langle{R}_{\rm EM/\mu}\rangle(B
   = 0)} 
\label{deltaB}
\end{equation}
 
As can be seen in the right panels of Fig.~\ref{BeffRatio}, the effect
of the geomagnetic field on $R_{\rm EM/\mu}$ is more important near the
core at all zenith angles. The reason for this is that only the
highest energy muons are not significantly deflected and stay close to
the core, and these are more likely to suffer hard interactions and
induce an electromagnetic shower. As a consequence $\rho_{\rm EM}$
increases and at the same time $\rho_{\mu}$ decreases because lower
energy muons are being deflected away from the core. These two effects
produce an overall increase in $R_{\rm EM/\mu}$. This increase is
small for $\theta \leq 78^{\circ}$ and $r>100$ m, with
$\Delta R_{B}\lesssim 20\%$. When $\theta = 80^{\circ}$ the effect
starts to be important in the case of maximal deviation ($\phi =
180^{\circ}$) where $\Delta R_{B} > 20\%$ for $r < 200$ m, whereas for
the case of minimal expected deviation ($\phi = 90^{\circ}$) the
relative difference remains smaller at all distances. At larger
angles $\theta>80^\circ$, the geomagnetic field has a strong influence
on $R_{\rm EM/\mu}$, even when $\phi \approx 90^{\circ}$ and the
effect is expected to be minimal. It can also be seen that the larger
the zenith angle, the farther from the shower core the influence of
the geomagnetic field is still important for the reasons explained
before.

In conclusion, the effect of the geomagnetic field on the ratio of the
electromagnetic to muonic number densities can be considered relevant
for showers at $\theta \gtrsim 84^{\circ}$. It should be noted that
the rate of events at such high zenith angles detected at ground level
by an array of detectors is small due to the reduced solid angle and
the small size of the projection of the array onto the direction of
the shower, so that this effect can be ignored for the purposes of
data analysis in a first approximation.

\section{Dependence of the ratio on primary energy, mass composition and hadronic model}

In this section we study the dependence of $R_{\rm EM/\mu}$ on primary
energy, mass of the primary cosmic ray initiating the shower, and
hadronic interaction model used to perform the simulations. These
dependences are studied neglecting the effect of the geomagnetic field
and averaging over azimuthal angle $\zeta$ in the shower plane.

\subsection{Energy dependence}

The more energetic a shower, its maximum occurs deeper in the
atmosphere, and therefore, the shower components are in a younger
stage of evolution. The lateral distribution of the electromagnetic
and muonic densities exhibits a characteristic behaviour with energy
and depth of shower maximum~\cite{Gaisser1990}. The lateral
distribution of the charged electromagnetic component due to the
cascading processes (mainly due to $\pi^0$ decay) is approximately
given by,

\begin{equation}
\ \rho_{e} =  N_{e} (E_{0},X-X_{max})\; f_{e}(r,X-X_{max})
\label{emcomp}
\end{equation}
where the total number of particles $N_e$ depends on primary energy
$E_0$ as $N_{e}\propto E_0^{\alpha}$. At shower maximum, $\alpha
\simeq 1$~\cite{Matthews:2005sd}. The lateral behaviour of the
electromagnetic component $f_{e}$ depends on shower age ($X-X_{max}$).

On the other hand, the lateral distribution of muons can be expressed as:

\begin{equation}
\ \rho_{\mu} =  N_{\mu}(E_{0},X-X_{max}) \; f_{\mu}(r) 
\label{emcomp}
\end{equation}
where the total number of muons $N_{\mu} \propto E_0^{\beta}$ (with $\beta <
1$) increases slower with the energy than the electromagnetic component
\cite{RaoEAS}. Also the lateral behaviour $f_{\mu}$ is approximately
independent of the shower age~\cite{Gaisser1990}.

\begin{figure*}
\begin{center}
\includegraphics[width=0.6\textwidth]{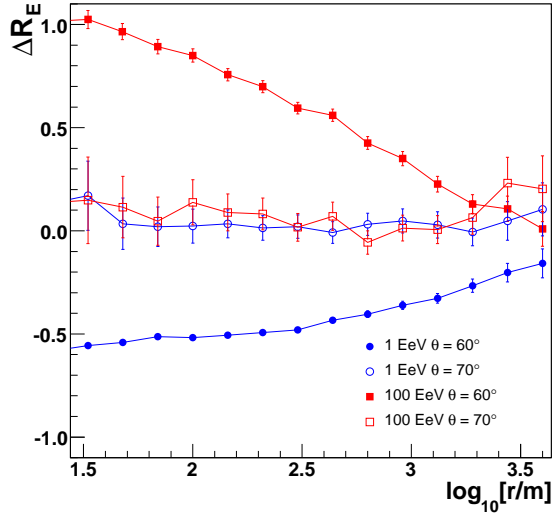}
\end{center}
\vspace{-15pt}
\caption{The relative difference $\Delta R_E$ between 
the ratio $R_{\rm EM/\mu}$ obtained in 1 EeV and 100 EeV proton-induced showers  
with respect to that obtained in 10 EeV proton shower
simulations. $\Delta R_E$ is shown as a function of distance to 
shower core $r$ at zenith angles $\theta=60^\circ$ and $\theta=70^\circ$.\label{Edep}}
\end{figure*}

Finally, as discussed before, the electromagnetic component due to
muon decay in flight is proportional to the muonic density. Hence, the
lateral distribution of this component is expected to have the same
energy dependence as the muonic one.

Combining all these facts, we expect the ratio $R_{\rm EM/\mu}=\rho_{\rm EM}
  /\rho_{\mu}$ to have a different behaviour depending on whether the
  electromagnetic component due to $\pi^0$ decay or the EM halo contributes
  more to the total electromagnetic density. We study the energy dependence
  of $R_{\rm EM/\mu}$ performing the relative difference between the ratio at
  a given energy with respect to that obtained for 10 EeV proton showers,
  $\langle{R}_{\rm EM/\mu}\rangle$ (see previous section):

\begin{equation} 
\Delta R_{E} = \frac{R_{\rm EM/\mu}(E)-\langle{R}_{\rm EM/\mu}\rangle({\rm 10~EeV})}
{\langle{R}_{\rm EM/\mu}\rangle({\rm 10~EeV})}
\label{deltaE}
\end{equation} 
In Fig.~\ref{Edep} we show $\Delta R_{E}$ for $E = $ 1 EeV and 100 EeV showers.
At $\theta = 70^{\circ}$, the energy dependence of the ratio is small
$\Delta R_{E}\lesssim 20\%$, because the electromagnetic density being
dominated by the contribution from the EM halo, is roughly proportional to the
muonic density, regardless of the shower energy. However for $\theta =
60^{\circ}$ there is a dependence of $\Delta R_E$ on the shower energy,
especially close to the shower core where the EM component due to cascading
processes contributes more to the EM density (see also Fig.~\ref{EMRemnant}).
In this angular range, we expect $R_{\rm EM/\mu}$ to behave as $\propto
E_0^{\alpha/\beta}$ and hence to depend on shower energy. As can be seen in
Fig.~\ref{Edep}, $\Delta R_E$ is larger (smaller) for 100 EeV (1 EeV) showers
than at 10 EeV because of the larger (smaller) EM component coming from the
cascading processes in the shower which penetrates more (less) in the
atmosphere.

\begin{figure*}
\begin{center}
\includegraphics[width=\textwidth]{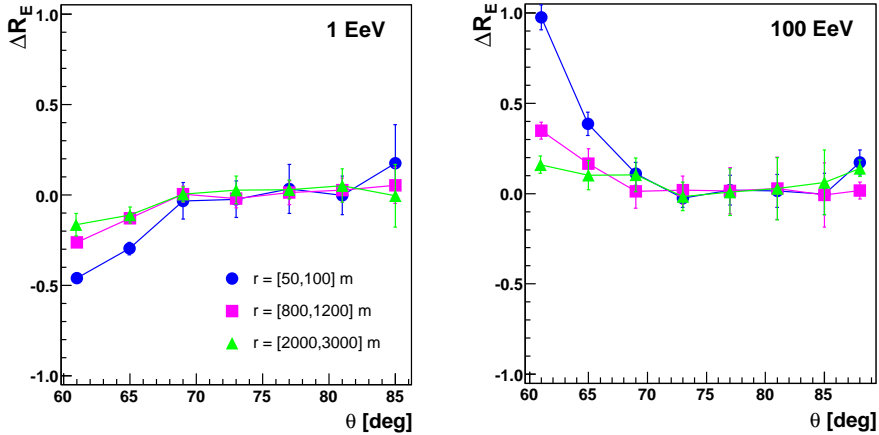}
\end{center}
\vspace{-15pt}
\caption{The relative difference $\Delta R_{E}$ between the ratio
    $R_{\rm EM/\mu}$ obtained in 1 EeV (left panel) and 100 EeV (right
    panel) proton-induced showers with respect to that obtained in 10
    EeV proton shower simulations. $\Delta R_E$ is plotted as a function
    of the shower zenith angle in different bins in distance to the
    core $r$.\label{Edepvsth}}
\end{figure*}

The dependence of $\Delta R_{E}$ on $\theta$ is studied in more detail in
Fig.~\ref{Edepvsth}, where we plot $\Delta R_{E}$ in different bins of $r$, as a
function of $\theta$ for 1 EeV and 100 EeV proton showers. Here, we confirm
what was said before, for zenith angles above $\sim 70^\circ$ the EM component
is only due to the EM halo and the ratio $R_{\rm EM/\mu}$ remains constant at
the same level with energy, while if $\theta\lesssim 70^\circ$ there is a
dependence on energy that increases as the distance to the core decreases.

\subsection{Primary mass dependence}

At present, the chemical composition of the cosmic rays at the highest
energies ($> 1$ EeV) remains uncertain. Some authors claim that cosmic
rays at these energies are mainly protons~\cite{Aloisio:2007rc}, and
others discuss the possibility of heavier elements such as iron
nuclei~\cite{Allard:2005cx}. The most recent results from extensive air shower experiments
do not allow to rise any firm
conclusion~\cite{Yamamoto:2007xj,Sokolsky:2008zza}. For this reason we
have studied the dependence of the ratio $\rho_{\rm EM} /\rho_{\mu}$
on the mass of the primary particle initiating the shower accounting
for protons and iron nuclei in our simulations.

Compared to a proton, an iron nucleus typically interacts higher in
the atmosphere, producing a shower with a smaller depth of
maximum. Also applying a simple superposition model an iron nuclei
produces a shower with $\sim 30\%-40\%$ more muons than a proton. As
a result, an iron-initiated shower is expected to have a smaller
electromagnetic density from cascading processes and a larger muonic
density than a proton shower, and therefore a smaller $R_{\rm EM/\mu}$
as long as the EM component due to cascading processes contributes
more than that due to the EM halo ($\theta<70^\circ$ and close to the
shower core). To demonstrate this, we have calculated the relative
difference between $R_{\rm EM/\mu}$ in iron showers at 10~EeV with
respect to that obtained in 10 EeV proton shower simulations

\begin{equation}
\Delta R_{\rm mass} = \frac{R_{\rm EM/\mu}(\rm{Fe}) - \langle{R}_{\rm EM/\mu}\rangle({\rm p})}
{\langle{R}_{\rm EM/\mu}\rangle({\rm p})} 
\label{deltaM}
\end{equation}
 
\begin{figure*}
\begin{center}
\includegraphics[width=0.6\textwidth]{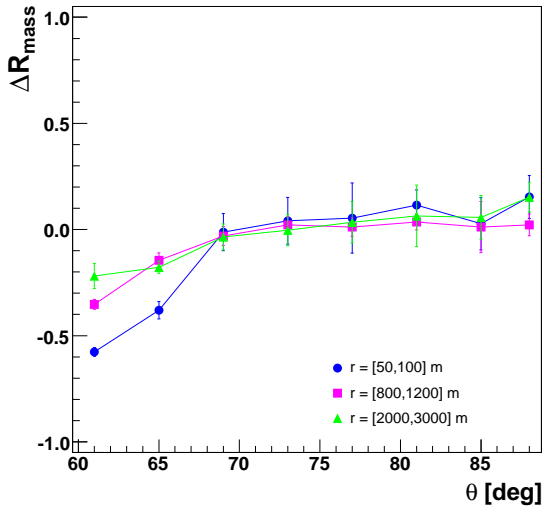}
\end{center}
\vspace{-15pt}
\caption{The relative difference $\Delta R_{\rm mass}$ between
    the ratio $R_{\rm EM/\mu}$ obtained in 10 EeV iron-induced shower
    simulations with respect to that obtained in 10 EeV proton shower
    simulations. $\Delta R_{\rm mass}$ is plotted as a function of the
    shower zenith angle in different bins in distance to the core
    $r$.\label{Massdepvsth}}
\end{figure*}

In Fig.~\ref{Massdepvsth} we show $\Delta R_{mass}$ as a function of
zenith angle in different bins of $r$. For reasons very similar to
those that explain the energy dependence of $R_{\rm EM/\mu}$ studied
before, we can explain the mass dependence.  For angles larger than
$\sim 70^{\circ}$ the dependence on the mass is negligible, because
only the EM halo is present and it is proportional to the muonic
density. Although the latter increases due to the larger mass of the
primary, the former increases accordingly and the ratio stays roughly
constant at the same level with mass. However when $\theta \lesssim
70^\circ$ we can see a clear dependence on mass, with iron showers
having smaller values of $R_{\rm EM/\mu}$ due to the fact that the EM
component is dominated by cascading processes and the iron-induced
shower penetrates less in the atmosphere.

\subsection{Hadronic model dependence}

At the highest energies, there is a lack of empirical knowledge about
the hadronic interactions which greatly influence shower development
\cite{Pierog:2006qu}. Laboratory experiments have studied particle
collisions only at centre-of-mass energies equivalent to fixed target
energies of $~ 10^{15}$ eV, so assumptions must still be made to
perform the interactions needed in the shower simulations at the
energies of interest ($> 10^{18}$ eV). This fact leads to
discrepancies between the different hadronic models on predictions of
inelastic cross-sections and inelasticity (multiplicity and energy of
the secondaries) (see for instance~\cite{Knapp:2002vs} for more
details). These quantities determine to a large extent the
longitudinal development of the air shower and, as a consequence, the
number densities of the EM and muonic components at ground.

\begin{figure*}
\begin{center}
\includegraphics[width=0.6\textwidth]{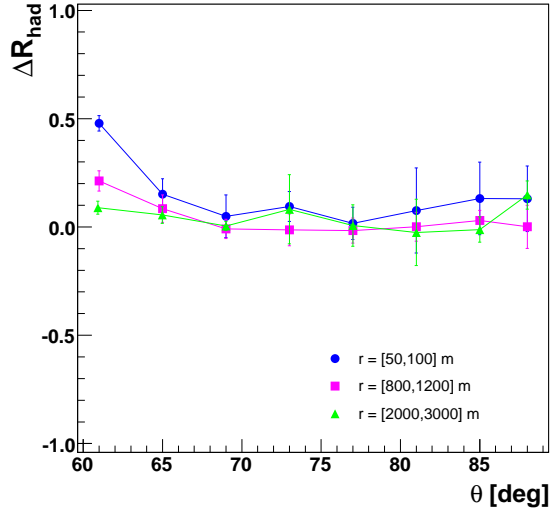}
\end{center}
\vspace{-15pt}
\caption{The relative difference $\Delta R_{\rm had}$ between the
    ratio $R_{\rm EM/\mu}$ obtained in 10 EeV proton-induced showers
    simulated with Sibyll 2.1, with respect to that obtained in 10 EeV
    proton showers simulated with QGSJET01. $\Delta R_{\rm had}$ is
    shown as a function of the shower zenith angle in different bins of distance
    to the shower core $r$.\label{Haddepvsth}}
\end{figure*}

In this work, we compare two high energy interaction models currently
used in cosmic ray physics: QGSJET01~\cite{qgsjet} and Sibyll
2.1~\cite{Engel:1999db}. For proton primaries at 10 EeV, the QGSJET
model predicts showers that on average develop higher in the
atmosphere and have $40\%$ more muons than showers simulated with
Sibyll. As a result, QGSJET predicts more muons at ground and a
smaller electromagnetic component due to cascading processes.

Following a similar line of reasoning as in the case of the energy and mass
dependence of the ratio of the EM to muonic densities, $R_{\rm EM/\mu}$ is
expected to be larger for showers simulated with Sibyll at zenith angles
$\lesssim 70^\circ$, and roughly independent on the model for $\theta$ larger
than $\sim 70^\circ$. This behaviour can be seen in Fig.~\ref{Haddepvsth},
where we plot the relative difference $\Delta R_{\rm had}$ between $R_{\rm
EM/\mu}$ for 10 EeV proton showers simulated with Sibyll 2.1 with respect to
the one obtained in showers simulated with QGSJET01:

\begin{equation}
\Delta R_{\rm had} = \frac{R_{\rm EM/\mu}(\rm{Sibyll}) - \langle{R}_{\rm EM/\mu}\rangle (\rm QGSJET)}
{\langle{R}_{\rm EM/\mu}\rangle (\rm QGSJET)} 
\label{deltaH}
\end{equation}

$\Delta R_{\rm had}$ is shown as a function of zenith angle in different bins of
$r$, to make the increasingly larger differences between Sibyll and QGSJET for
small distances to the shower axis more apparent, as expected from the
dominance of the EM component due to cascading processes near the core.

\begin{figure*}
\begin{center}
\includegraphics[width=\textwidth]{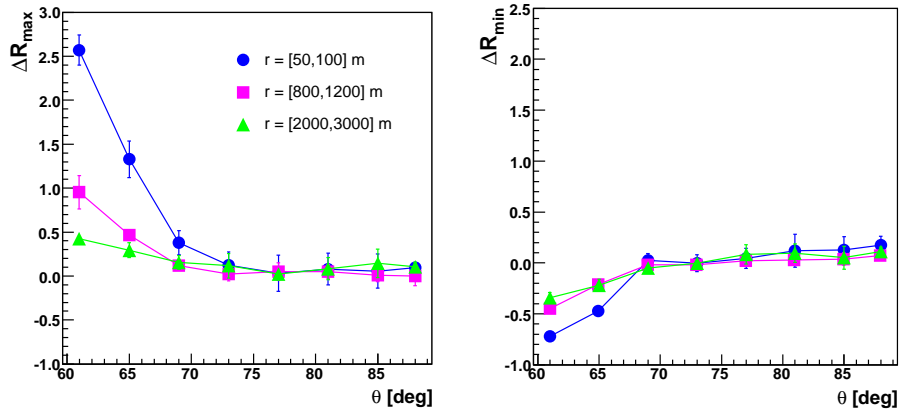}
\end{center}
\vspace{-15pt}
\caption{The maximum $\Delta R_{\rm max}$ (left panel) and
    minimum $\Delta R_{\rm min}$ (right panel) relative differences in
    the ratio $R_{\rm EM/\mu}$ obtained respectively for 100 EeV
    proton showers simulated with Sibyll 2.1 and 1 EeV iron showers
    simulated with QGSJET01, with respect to proton showers at 10 EeV
    energy simulated with QGSJET01. $\Delta R_{\rm max}$ and
    $\Delta R_{\rm min}$ are both shown as a function of the shower
    zenith angle in different bins in distance to the shower core $r$.\label{Largestdepvsth}}
\end{figure*}

\subsection{Summary and discussion}

Summarizing this section, we have studied the effect of energy, mass
composition and hadronic interaction model on the ratio of
electromagnetic to muonic densities in absence of geomagnetic field
effect, using the ratio obtained for 10~EeV proton showers simulated
with QGSJET01 as reference. Combining all these dependences, we find
that the extreme differences with respect to the reference ratio
correspond to:

\begin{itemize}

\item Maximum: 100 EeV proton showers simulated with Sibyll 2.1.

\item Minimum: 1 EeV iron showers simulated with QGSJET01.

\end{itemize}

In Fig.~\ref{Largestdepvsth} we show the relative differences between
both cases and the reference ratio ($\Delta R_{\rm max}$ and
$\Delta R_{\rm min}$) as a function of $\theta$ in different bins in
$r$. The figure illustrates the extreme cases in the dependency of
$R_{\rm EM/\mu}$ on energy, mass and model one should expect.


\section{Fluctuations of the particle densities}
\label{sec:fluctuations}

In this section, we attempt to estimate the effect of physical
fluctuations on the muon and electromagnetic number densities using
simulations. The particle densities in individual showers are
affected by different sources of fluctuation, namely:
\begin{itemize}

\item Physical (intrinsic) ``shower-to-shower'' fluctuations due to
  fluctuations in the atmospheric depth of the first interactions in
  the shower, fluctuations in secondary particle production, etc., in
  general fluctuations in the cascading processes during the
  development of the shower.

\item ``Artificial'' fluctuations due to the thinning procedure in the
  shower simulation.

\end{itemize}
The ideal way of computing the ``shower-to-shower'' fluctuations of
the electromagnetic and muonic components is to perform full
(non-thinned) shower simulations. This is unfortunately not feasible
due to the large computing time and huge disk space needed to store
the information on particles at ground at the highest energies (even
for a single shower). One has to rely on thinned simulations in which
artificial fluctuations are introduced. To overcome this problem, we
have devised a simple approach based on a method given in
\cite{Risse:2002yd} to estimate how physical fluctuations affect the
electromagnetic and muonic particle distributions when obtained from
tracked particles with weights $w_i$. Here, we neglect the azimuthal
asymmetry of the densities and obtain the densities in concentric
rings in $r$ around the shower core of area $A_{\rm r}$.

The particle density corresponding to $n$ particles with individual
weights $w_i$ falling in an area $A_{\rm r}$ is calculated as:

\begin{equation}
\ \rho =  \sum_{i=1}^{n} \frac{w_{i}}{A_{\rm r}} = \frac{N}{A_{\rm r}}
\label{density}
\end{equation}
 so that the fluctuations of the particle densities stem from the
 fluctuations in the total particle number after accounting for the
 weights: $\sigma_{\rho} \approx \sigma_{N}$. Defining the average
 weight of the $n$ particles falling in a ring of area $A_{\rm r}$ as
 $\overline{w} = \sum_{i=1}^{n} w_{i} / n$, the unweighted particle
 number can be approximated as $N\simeq \overline{w}n$ and the
 standard deviation of $N$ can be expressed as:

\begin{equation}
\ \sigma_{N}^2 = \langle N^2 \rangle -\langle N\rangle^2 = \langle \overline{w}^2
n^2 \rangle- \langle \overline{w} n \rangle^2
\label{sigmaN}
\end{equation}

Assuming that the mean particle weight does not fluctuate from shower
to shower as in the ideal case of an unthinned shower, and taking it
to be equal to the average weight over all the particles followed
explicitly in the shower simulation, we obtain:

\begin{equation}
\ \sigma_{N}^2 =  \overline{w}^2 \, (\langle n^2 \rangle - \langle  n \rangle^2) = \overline{w}^2 \sigma_{n}^2
\label{sigmaN_shtosh}
\end{equation}
 
Eq.~\ref{sigmaN_shtosh} represents an approximate way of obtaining the
intrinsic physical shower-to-shower fluctuations of the muonic and
electromagnetic particle densities from Monte Carlo simulations, which
gives a good account of the physical fluctuations for thinning levels
smaller than $\sim 10^{-6}$ \cite{hansen_fluct}. In fact, in the case
of an unthinned shower, in which only the intrinsic fluctuations
should be present, $\bar w=1$, $N=n$ and Eq. 8 gives
$\sigma_N^2=\sigma_n^2$ as expected.

\begin{figure*}
\begin{center}
\includegraphics[width=0.49\textwidth]{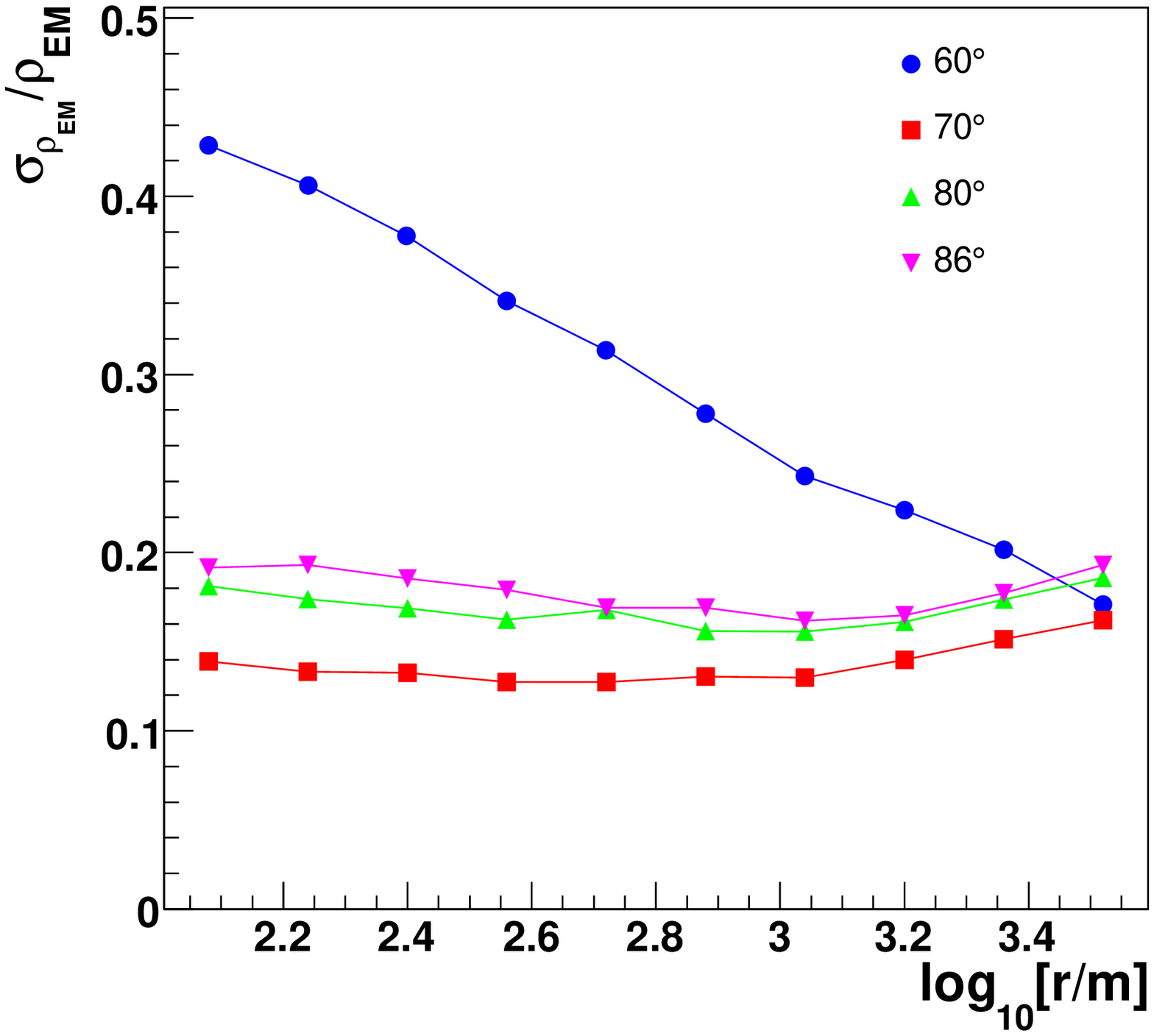}
\includegraphics[width=0.49\textwidth]{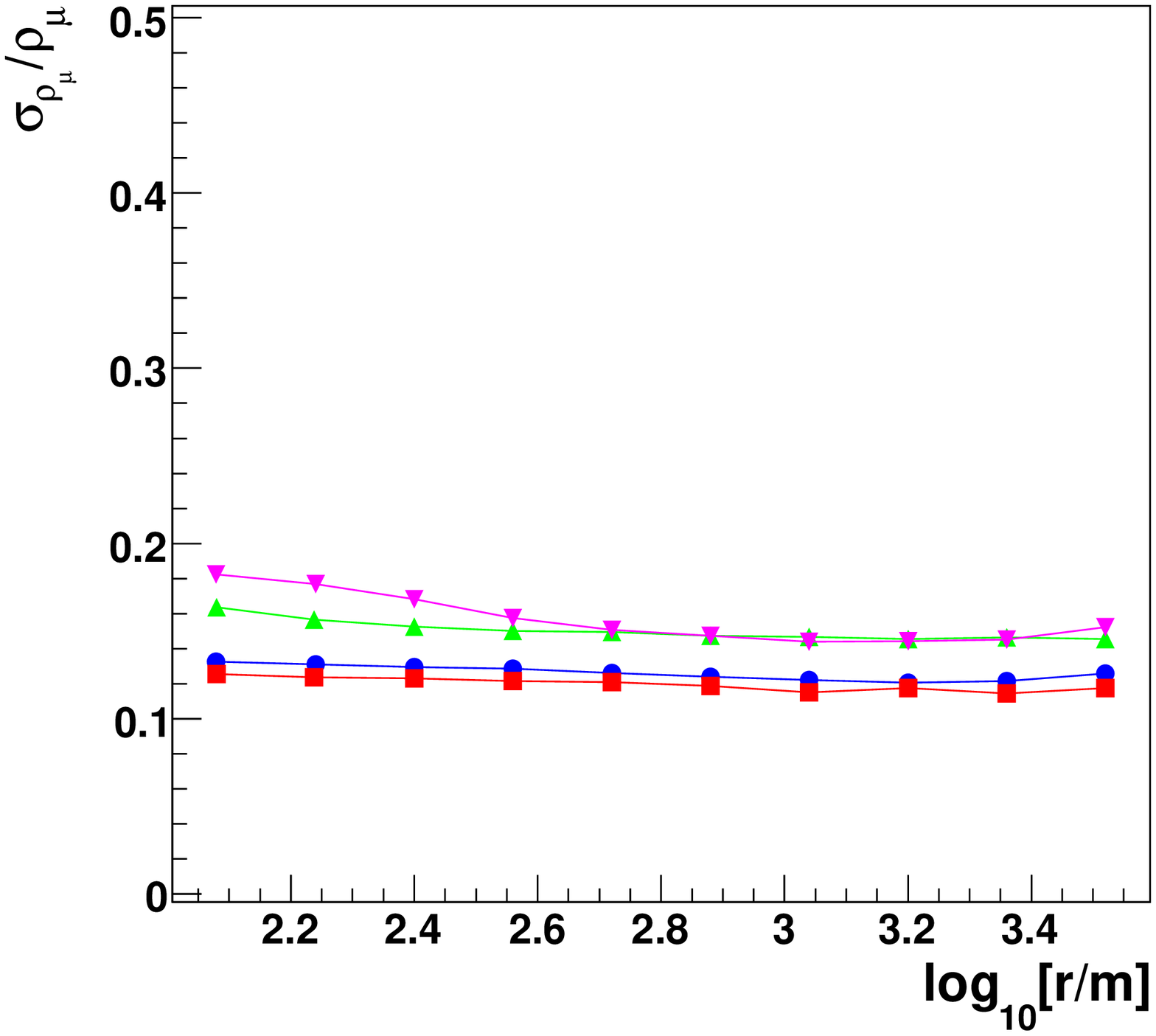}
\end{center}
\vspace{-15pt}
\caption{Estimated relative physical fluctuations of the
    electromagnetic (left panel) and muonic (right panel) densities at
    ground level as a function of the distance to the shower core in
    the shower plane, obtained in 10 EeV proton showers
    simulated with AIRES + QGSJET01 and with a thinning level of $10^{-7}$.\label{Signalfluct}}
\end{figure*}

In Fig.~\ref{Signalfluct} we show the relative physical fluctuations
($\sigma^{\rm rel} = \sigma_{\rho}/\rho$)
of the electromagnetic (top) and muon (bottom) densities as a function
of the distance to the shower core for 10 EeV proton showers at
different zenith angles. These have been computed obtaining first
$\sigma_{n}^2$ and $\overline{w}^2$ from the simulations and then
applying Eq.~\ref{sigmaN_shtosh}. The fluctuations in the muonic
component are roughly independent of $r$ in the wide range of
distances to the core plotted in the figure, and remain at the level
of $\lesssim 20\%$.  On the other hand, the fluctuations in the
electromagnetic component seem to depend on the contribution of the EM
halo. At $\theta = 60^{\circ}$ the decrease of the relative
fluctuations as the distance to the core increases could be
interpreted as the EM halo is becoming the dominant contribution to
the EM density and mimicking the behaviour of the fluctuations in the
muonic component. Similarly, for $\theta \geq 70^{\circ}$ the
fluctuations mimic the behaviour of those of the muonic component,
they are roughly independent of $r$ and remain at the level of
$\lesssim 20\%$.

\section{Summary and conclusions}

We have studied the characteristics of the particle densities of the
electromagnetic and muon components of inclined showers at the ground
level using Monte Carlo simulations. We have shown that the
electromagnetic component is composed of several sub-components
originated in cascading and muonic processes. These different
contributions to the electromagnetic component differ from each other
on their behaviour with distance to the core, zenith angle and angular
position ($\zeta$).

 We have studied the ratio of the electromagnetic to muon densities
 ($R_{\rm EM/\mu}$) as a function of several parameters. Firstly, we
 have characterised the dependence of this ratio on the distance to
 the core and shower zenith angle (see Fig.~\ref{EmMuRatio}). Near the
 core up to $\theta \sim 70^{\circ}$ its behaviour is explained by the
 increasing absorption of the contribution to the EM component due to
 $\pi^0$ decay and beyond $70^{\circ}$ by the dominance of the
 contribution to the EM component due hard muon processes. Far from
 the core ($> 1$ km), the ratio is compatible with an almost constant
 value because the electromagnetic component due to muon decay in
 flight dominates the electromagnetic density at ground. Then, we have
 studied the dependence of $R_{\rm EM/\mu}$ on the azimuthal position
 $\zeta$. For showers with $60^{\circ} < \theta < 70^{\circ}$ we have
 found an azimuthal asymmetry, which is mostly due to the longitudinal
 development effect. Moreover, we have studied the effect of the Earth
 magnetic field in this ratio finding that is important for showers at
 $\theta \gtrsim 84^{\circ}$. 

Considering all these dependences above and neglecting the geomagnetic
field effect, we propose a parameterisation of the $R_{\rm
  EM/\mu}(r,\theta,\zeta)$ that predicts the simulated data to a good
precision level (see Fig.~\ref{ExampleAsyEmMuRatio}). The fits are
valid for 10 EeV proton showers simulated with the hadronic model
QGSJET01, and in the ranges $\theta \in [60^{\circ},89^{\circ}]$ and
$r \in [20,4000]$ m.

We have characterised the dependence of this ratio with the primary
energy, the primary mass composition and the hadronic interaction
model used in the simulations. The general result is that at zenith
angles $ \gtrsim 70^{\circ}$ the ratio remains constant because only
the electromagnetic halo contributes to the electromagnetic component.

Finally, we have estimated the effect of physical fluctuations on the
electromagnetic and muon densities. The muon fluctuations remain
roughly constant at the level of $ \lesssim 20\%$. While, the EM
fluctuations depend on the dominance of the electromagnetic halo on
the EM component, and tend to be constant $\lesssim 20\%$ for higher
zenith angles.

In Table~\ref{tabla1} we summarise some results for two different
distances to the core and different shower zenith angles. Note that in
the angular region $60^{\circ} < \theta < 70^{\circ}$ the presence of
the remnant of the electromagnetic shower due to cascading processes
from $\pi^0$ decay of hadronic origin entails important dependences of
the ratio on the energy, mass composition and hadronic
model. Therefore, this region could be the most suitable one for
studies about composition and hadronic interaction models by means of
the electromagnetic component in the shower at ground level. However
in the region $\theta \gtrsim 70^{\circ}$ the ratio is not sensitive
to composition and hadronic model, and this angular region might 
be in principle more suitable to study the UHECR energy spectrum with
inclined events detected by ground detectors.

\begin{table}[]
\begin{center}
\begin{tabular}{||c|c|c|c|c|c|c|c||}
\hline\hline
\small{r (m)} & \small{$\theta$ (deg)} & \small{$\Delta R_{{B}_{\phi = 180^{\circ}}}$} & \small{$\Delta R_{E=100 EeV}$} & \small{$\Delta R_{\rm mass}$} & \small{$\Delta R_{\rm had}$} & \small{$\sigma_{{\rho_{\rm EM}}}^{\rm rel}$}& \small{$\sigma_{\rho_{\mu}}^{\rm rel}$}\\
\hline
\hline 100 &  60 & -2  & 83 &  -61 & 40 &  45 & 13  \\
\cline{2-8} & 70 & 7  & 6 &  5 & 11 &  14 &  12  \\
\cline{2-8} & 80 & 29  & -2 &  5 & 1 &  17 &  16  \\
\hline 1000 &  60 & -2  & 33 &  -42 & 19 &  26 &  13  \\
\cline{2-8} & 70 & 7    & 4 &  2 & 3 & 14 & 12  \\
\cline{2-8} & 80 & 9  & -2 &  -2 & -1 & 17&  15  \\
\hline\hline
\end{tabular}
\caption{Summary of the dependences of the ratio of EM to muonic component 
on the geomagnetic field, energy, primary mass and hadronic model, using protons 
at $E=10^{19}$ eV simulated with the QGSJET01 model and with no geomagnetic field 
as reference (see Eqs.~\ref{deltaB}, \ref{deltaE}, \ref{deltaM} and \ref{deltaH} for the 
definitions of the different $\Delta R$). The dependences are given at 
different distances from the shower axis 100 m and 1 km, and
  different zenith angles $60^{\circ}$, $70^{\circ}$ and $80^{\circ}$. 
We also give the relative physical fluctuations of the EM and muonic 
components as obtained with Eq.~\ref{sigmaN_shtosh}.
The results are given in percentage $\%$.\label{tabla1}}
\end{center}
\end{table}

\section{Acknowledgments}

The authors thank Gonzalo Rodr\'iguez-Fern\'andez and our colleagues of the
Pierre Auger Collaboration for support and discussions on this topic. The
Helmholtz association, Germany (HHNG-128 grant), Ministerio de Ciencia e
Innovaci\'on (FPA 2007-65114), the Spanish Consolider-Ingenio 2010 Programme
CPAN (CSD2007-00042), Xunta de Galicia (PGIDIT 06 PXIB 206184 PR) and
Conseller\'\i a de Educaci\'on (Grupos de Referencia Competitivos --
Consolider Xunta de Galicia 2006/51), and FEDER Funds, Spain, are acknowledged
for providing funding.  In\'es Vali\~no gratefully acknowledges the financial
support from ``Fundaci\'on Pedro Barri\'e de la Maza'' (Spain). The authors
also thank CESGA (Centro de Supercomputaci\'on de Galicia) for computing
resources.

\appendix

\section{Parameterisation of the muonic and electromagnetic particle densities}

We give here the parameterisations performed for the lateral 
distributions of the electromagnetic and muonic densities at ground level
projected onto the shower plane. The densities are given as a function 
of distance to the core $r$ in the shower plane and shower zenith angle $\theta$,
averaging over the azimuthal angle $\zeta$. In Appendix B we also give a
parameterisation of the azimuthal asymmetry.  

The fits are valid for 10 EeV proton showers simulated with the hadronic
model QGSJET01, and in the ranges $\theta \in [60^{\circ},89^{\circ}]$ and $r \in [20,4000]$ m.

\subsection{Muonic component}

We have used two different functional forms to account for the
different behaviour of the lateral distributions at large distances
from the core. Near the core, we use a form based on the Vernov
functional form~\cite{Vernov:1968} and for larger distances a function
based on the Nishimura-Kamata-Greisen functional
form~\cite{Kamata:1958,Greisen:1960wc}:
 
\begin{equation}
 \rho_{\mu}(r) =  \left\{ \begin{array}{ll}
   A'\; r^{-B'}\exp( -\frac{r}{C'})& \;\;\; \mbox{for $r < r_{0}$}\\
    A \left(\frac{r}{B}\right)^{-C} \left( 1 + \frac{r}{B}\right)^{-D} + E & \;\;\;  \mbox{for $r \geq r_{0}$}
 \end{array} \right.
\label{MuSignals_fit}
\end{equation}
with $r_{0} = 250$ m. All the parameters, except $B = 2000$ m and $D =
4.64$, depend on $\theta$.

For distances $r > r_{0}$, the parameters can be fitted with:
\begin{equation}
\hspace{-45pt} A  = 0.221 + 6.490  \cos\theta \,- 14.290 \cos^2\theta +   61.693 \cos^3\theta  - \, 66.426 \cos^4\theta 
\nonumber
\end{equation}
\begin{equation}
\hspace{-122pt} C  = 0.460 + 2.646 \cos\theta - 5.550 \cos^2\theta + 4.927 \cos^3\theta 
\nonumber
\end{equation}
\begin{equation}
\begin{split}
\hspace{-2pt} E = 10^{-4} \times \, \left\{ \begin{array}{ll}  ( - 4.2 - 23.0 \cos\theta + 174.8 \cos^2\theta - 419.9 \cos^3\theta) \hspace{25pt}  \mbox{for $\theta < 76^{\circ}$} \hspace{2.9cm}
\\  ( - 4.8  -  0.3\cos\theta - 9.1 \cos^2\theta) \hspace{102pt} \mbox{for $\theta \geq 76^{\circ}$}
\end{array} \right.
\end{split}
\nonumber
\end{equation}

For distances $r < r_{0}$, the parameters can be fitted with:
\begin{equation}
\hspace{-245pt} A'  = \frac{\rho_{\mu}(r_0)} {r_{0}^{-B'}\exp( - r_0 / C' )}
\nonumber
\end{equation}
\begin{equation}
\hspace{-45pt} B'  = 0.438 + 1.978 \cos\theta - 11.593 \cos^2\theta + 29.724 \cos^3\theta  - \, 24.553 \cos^4\theta
\nonumber\end{equation}
\begin{equation}
\hspace{-70pt} C'  = 417.594 -953.421 \cos\theta + 1960.080 \cos^2\theta -1342.210 \,\cos^3\theta 
\nonumber
\end{equation}

\subsection{Electromagnetic component}

We have fitted the lateral distribution of the electromagnetic
component to the same functional forms used for the muonic component
(see Eq.~\ref{MuSignals_fit}) and in the same radial ranges. In this
case, the parameters show a different dependence on the zenith angle
below and above $\theta = 70^{\circ}$, reflecting the dominance of the
contribution of the EM halo to the total EM density.

For distances $r > r_{0}$, \, $B = 1200$ m and the remaining parameters can be fitted with:
\begin{equation}
A(\theta ) =  \left\{ \begin{array}{ll}
\exp(-13.120 + 33.372 \cos\theta) + 30.497 - 44.807 \cos\theta \hspace{29pt} \mbox{for $\theta < 70^{\circ}$}\\
8.518  - 39.824 \cos\theta + 294.769 \cos^2\theta -350.548 \cos^3\theta \hspace{24pt} \mbox{for $\theta \geq 70^{\circ}$}
 \end{array} \right. 
\nonumber
\end{equation}
\begin{equation}
 C(\theta ) =  \left\{ \begin{array}{ll}
 22.228 - 167.812 \cos\theta +  423.455 \cos^2\theta -340.857 \cos^3\theta \hspace{14pt} \mbox{for $\theta < 70^{\circ}$}\\
  0.512 + 1.051 \cos\theta -  1.207\cos^2\theta \hspace{110pt} \mbox{for $\theta \geq 70^{\circ}$}
 \end{array} \right.
 \nonumber
\end{equation}
 \begin{equation}
D(\theta ) =  \left\{ \begin{array}{ll}
 -14.057 +  176.968 \cos\theta  - 531.461\cos^2\theta + 494.975  \cos^3\theta \hspace{6pt} \mbox{for $\theta < 70^{\circ}$}\\
  4.104 \hspace{228pt} \mbox{for $\theta \geq 70^{\circ}$}
 \end{array} \right. 
\nonumber
\end{equation}
 \begin{equation}
 E(\theta ) =  \left\{ \begin{array}{ll}
 0.020 - 0.084 \cos\theta + 0.069\cos^2\theta \hspace{112pt} \mbox{for $\theta < 70^{\circ}$}\\
 -0.001 \hspace{222pt} \mbox{for $\theta \geq 70^{\circ}$}
 \end{array} \right. 
\nonumber
\end{equation}

For distances $r < r_{0}$ , $C' = 284$ m  and the remaining parameters can be fitted with: 
\begin{equation}
\hspace{-230pt} A' (\theta) = \frac{\rho_ {EM}(r_0)} {r_{0}^{-B'}\exp( - r_0 / C' )} 
\nonumber
\end{equation}
\begin{equation}
B' (\theta ) =  \left\{ \begin{array}{ll}
 0.500 + 0.580 \cos\theta - 0.848 \cos^2\theta \hspace{110pt} \mbox{for $\theta < 70^{\circ}$}\\
 13.642 - 104.697 \cos\theta + 268.581\cos^2\theta - 216.302 \cos^3\theta \hspace{14pt} \mbox{for $\theta \geq 70^{\circ}$}
 \end{array} \right. 
\nonumber
\end{equation}

\section{Parameterisation of the azimuthal asymmetry in the ratio of the electromagnetic to muonic lateral densities}

We have also parameterised the asymmetry parameter $\Delta_\zeta$. 
The following fits are valid for 10 EeV proton showers simulated with 
QGSJET01, in the ranges $\theta \in [60^{\circ},68^{\circ}]$,
$\zeta \in [-180^{\circ},180^{\circ}]$ and $r \in [20,4000]$ m.
Note that the asymmetry is only important for $\theta\lesssim70^\circ$.

\begin{equation}
\ \Delta_{\zeta} = \alpha \;(r \, - \, r_0 ) + \beta \; (log^{2}_{10}r \, - \, log_{10}^{2}r_0)
\label{AsymParam}
\end{equation}
where $r_0 = 20$ m. 

The parameters $\alpha$ and $\beta$ can be parameterised using a Cauchy-type function modified: 
\[ \alpha  =  - \alpha_1 \left[ \frac{1}{\pi} \left( \frac{\alpha_2}{\zeta^2 + \alpha_2^2} \right) - \alpha_3\right]  \hspace{1.cm} \beta =  \beta_1 \left[ \frac{1}{\pi} \left( \frac{\beta_2}{\zeta^2 + \beta_2^2} \right) - \beta_3\right] \]
with $\zeta \in [-180^{\circ},180^{\circ}]$ and:
\[ \hspace{-90pt} \alpha_1 = - 441.6080 + 20.7691 \, \theta - 0.324046\,\theta^2 + 0.00167832\, \theta^3\]
\[ \hspace{-250pt} \alpha_2 = 589.326 - 7.10\, \theta \]
\[ \hspace{-282pt} \alpha_3 =  0.0013437 \]
\[ \hspace{-110pt} \beta_1 = -33332.80 +  1547.31 \,\theta  -23.7198\,\theta^2 + 0.120283\,\theta^3\]
\[ \hspace{-204pt} \beta_2 =  265.469  - 5.400\, \theta + 0.040\, \theta^2\]
\[ \hspace{-170pt} \beta_3 = 10^{-5}\, (6318.27 -196.80\, \theta +  1.58\,\theta^2) \]



\begin{thebibliography}{99}


\bibitem{Ave:1999cp}
M.~Ave, et~al.,~ Proc. 26$^{\rm th}$ Int. Cosmic Ray Conference, Salt Lake
  City, 1 (1999), p. 365.

\bibitem{Yoshida:2001pw}
S.~Yoshida, et~al.,~ Proc. 27$^{\rm th}$ Int. Cosmic Ray Conference, Hamburg,
  (2001), p. 1142.

\bibitem{Abraham:2004dt}
J.~Abraham, et~al., Nucl. Instrum. Meth. A523 (2004) 50--95.

\bibitem{TA}
{h}ttp://www.telescopearray.org.

\bibitem{Kawai:2008zza}
H.~Kawai, et~al., Nucl. Phys. Proc. Suppl. 175-176 (2008) 221--226.

\bibitem{FacalSanLuis:2007it}
P.~Facal San~Luis~[Pierre Auger Collaboration], Proc. 30$^{\rm th}$ Int. Cosmic
  Ray Conference, Merida, 4 (2007), p. 339.

\bibitem{Inoue:1999cn}
N.~Inoue~[AGASA Collaboration], Proc. 26$^{\rm th}$ Int. Cosmic Ray Conference,
  Salt Lake City, 1 (1999), p. 357.

\bibitem{Ave:2000xs}
M.~Ave, R.~A. Vazquez, E.~Zas, Astropart. Phys. 14 (2000) 91.

\bibitem{Newton:2007qi}
D.~Newton~[Pierre Auger Collaboration], Proc. 30$^{\rm th}$ Int. Cosmic Ray
  Conference, Merida, 4 (2007), p. 323.

\bibitem{Berezinsky1975}
V.~Berezinsky, A.~Smirnov, Astrophys. Space Sci. 32 (1975) 461.

\bibitem{Cillis:2000xc}
A.~N. Cillis, S.~J. Sciutto, Phys. Rev. D64 (2001) 013010.

\bibitem{aires}
S.~Sciutto, http://www.fisica.unlp.edu.ar/auger/aires/.

\bibitem{qgsjet}
N.~Kalmykov, S.~S. Ostapchenko, A.~I. Pavlov, Nucl. Phys. Proc. Suppl. 52B
  (1997) 17--28.

\bibitem{Engel:1999db}
R.~Engel, et~al.,~Proc. 26$^{\rm th}$ Int. Cosmic Ray Conference, Salt Lake
  City, 1 (1999), p. 415.

\bibitem{Linsley}
J. Linsley, private communication by M. Hillas (1988).

\bibitem{Hillas:1997tf}
A.~M. Hillas, Nucl. Phys. Proc. Suppl. 52B (1997) 29--42.

\bibitem{Billoir:2008zz}
P.~Billoir, Astropart. Phys. 30 (2008) 270--285.

\bibitem{Ave:2000dd}
M.~Ave, R.~A. Vazquez, E.~Zas, J.~A. Hinton, A.~A. Watson, Astropart. Phys. 14
  (2000) 109--120.

\bibitem{AveThesis}
M.~D. Ave~Pernas, High {E}nergy {A}ir {S}howers, Ph{D} {T}hesis, Univ. de
  Santiago de Compostela, Spain (1999).

\bibitem{Hillas:1969zza}
A.~M. Hillas, et~al.,~Proc. 11$^{\rm th}$ Int. Cosmic Ray Conference, Budapest,
  3 (1969), p. 533.

\bibitem{IGRF}
Http://www.ngdc.noaa.gov/IAGA/vmod/igrf.html.

\bibitem{Dova:2001jy}
M.~T. Dova, L.~N. Epele, A.~G. Mariazzi, Astropart. Phys. 18 (2003) 351--365.

\bibitem{Gaisser1990}
T.~K. Gaisser, Cosmic {R}ays and {P}article {P}hysics, Cambridge University
  Press, New York, USA, 1990.

\bibitem{Matthews:2005sd}
J.~Matthews, Astropart. Phys. 22 (2005) 387--397.

\bibitem{RaoEAS}
M.~V.~S. Rao, B.~V. Sreekantan, Extensive {A}ir {S}howers, World Scientific
  Singapore, Singapore, 1999.

\bibitem{Aloisio:2007rc}
R.~Aloisio, V.~Berezinsky, P.~Blasi, S.~Ostapchenko, {Signatures of the
  transition from galactic to extragalactic cosmic rays}, Phys. Rev. D77 (2008)
  025007.

\bibitem{Allard:2005cx}
D.~Allard, E.~Parizot, A.~V. Olinto, {On the transition from Galactic to
  extragalactic cosmic- rays: spectral and composition features from two
  opposite scenarios}, Astropart. Phys. 27 (2007) 61--75.

\bibitem{Yamamoto:2007xj}
T.~Yamamoto~[Pierre Auger Collaboration], Proc. 30$^{\rm th}$ Int. Cosmic Ray
  Conference, Merida, 4 (2007), p. 335; J. Bellido~[Pierre Auger
  Collaboration], Proc. 31$^{\rm th}$ Int. Cosmic Ray Conference, Lodz (2009),
  astro-ph/0906.2319.

\bibitem{Sokolsky:2008zza}
P.~Sokolsky, Nucl. Phys. Proc. Suppl. 175-176 (2008) 207--212.

\bibitem{Pierog:2006qu}
T.~Pierog, R.~Engel, D.~Heck, Czech. J. Phys. 56 (2006) A161--A172.

\bibitem{Knapp:2002vs}
J.~Knapp, D.~Heck, S.~J. Sciutto, M.~T. Dova, M.~Risse, Astropart. Phys. 19
  (2003) 77--99.

\bibitem{Risse:2002yd}
M.~Risse, et~al.,~Proc. 27$^{\rm th}$ Int. Cosmic Ray Conference, Hamburg,
  (2001), p. 522.

\bibitem{hansen_fluct}
P.~M. Hansen, et~al.,~Proc. 31$^{\rm th}$ Int. Cosmic Ray Conference, Lodz
  (2009), ID 0167.

\bibitem{Vernov:1968}
S.~N. Vernov, et~al., Can. J. Phys. 46 (1968) S197.

\bibitem{Kamata:1958}
K.~Kamata, J.~Nishimura, Progr. Theor. Phys. Suppl. 6 (1958) 93.

\bibitem{Greisen:1960wc}
K.~Greisen, Ann. Rev. Nucl. Part. Sci. 10 (1960) 63--108.

\end{thebibliography}
\end{document}